\documentclass[11pt,a4paper]{article}
\usepackage{graphicx}
\usepackage{balance}  
\usepackage{wrapfig}
\usepackage{microtype}
\usepackage{hyperref}
\usepackage{verbatim}
\usepackage[]{algorithm2e}
\SetAlFnt{\small}
\usepackage{algorithmic}
\algsetup{linenosize=\tiny}

\usepackage{booktabs} 
\usepackage{xspace}
\usepackage{amsmath}
\usepackage{amssymb}
\usepackage{tikz-cd}
\usepackage{subcaption}
\usepackage{enumitem}
\setlist{nolistsep}

\usepackage{listings,textcomp}
\usepackage{color}
\definecolor{Orange}{HTML}{CF4A30}
\definecolor{gray}{rgb}{0.4,0.4,0.4}
\definecolor{darkblue}{rgb}{0.0,0.0,0.6}
\definecolor{darkred}{rgb}{0.45,0,0}
\definecolor{darkgreen}{rgb}{0,0.30,0.20}
\definecolor{darkpurple}{RGB}{120, 0, 180}
\colorlet{keywordcolor}{darkblue}
\newcommand{\kwfont}{\color{keywordcolor}\bfseries}

\colorlet{labelcolor}{darkgreen}
\newcommand{\lblfont}{\color{labelcolor}}

\colorlet{keycolor}{darkred}
\newcommand{\keyfont}{\color{keycolor}}

\colorlet{structcolor}{black}
\newcommand{\structfont}{\color{structcolor}\bfseries}

\newenvironment{pseudoquery}[1][htb]
  {
   \begin{algorithm}[#1]%
  }{\end{algorithm}}

\lstdefinelanguage{cypher}
{
  columns=fullflexible,
  otherkeywords={<,>,-,[,],(,),\{,\},|,.,=,:},
  keywords=[1]{},
  keywordstyle=,
  keywordstyle=[1]\color{darkpurple},
  keywords=[2]{MATCH,WHERE,WITH,OPTIONAL,RETURN,MERGE,CREATE,SET,DELETE,
  REMOVE,CALL,YIELD,AS,ORDER,BY,DESC,LIMIT,UNWIND,
  STRING,INTEGER,TIMESTAMP, FOREACH, IN, IS, NULL, DATE,
  GRAPH,TYPE},
  keywordstyle=[2]\kwfont,
  keywords=[3]{<,>,-,[,],(,)},
  keywordstyle=[3]\structfont,
  keywords=[4]{\{,\},|,.,=,:},
  keywordstyle=[4]\structfont,
  keywords=[5]{Student,Person,Message,Comment,Post, Place, Continent, Country, City, Tag, University, Organisation, Forum, TagClass, Company, MessageHeader, MessageBody},
  keywordstyle=[5]\lblfont,
  keywords=[6]{STUDIES_AT,REPLY_OF, KNOWS, LIKES, HAS_CREATOR,
  	STUDY_AT, WORK_AT, IS_LOCATED_IN, HAS_TYPE, CONTAINER_OF, HAS_TAG, IS_PART_OF, HAS_INTEREST, HAS_MEMBER, HAS_MODERATOR, IS_SUBCLASS_OF},
  keywordstyle=[6]\lblfont,
  keywords=[7]{collect,count,size,exists,id,toInteger,round,rand,filter,
  name,age,studid,
gender, birthday, email, speaks, url, title, joinDate, workFrom, classYear,  creationDate,browserUsed,locationIP,content,length,language,imageFile,firstName,lastName},
  keywordstyle=[7]\keyfont,
  string=[m]{"},
  stringstyle=,
  moredelim=*[is][\bfseries]{**}{**},
  moredelim=*[is][\fixeddepthunderline]{__}{__},
}
\lstnewenvironment{cypher}[1][]{\lstset{language=cypher,#1}}{}
\newcommand{\cypherinline}[1]{\lstinline$#1$}

\newcommand{\set}[1]{\ensuremath{\left\{#1\right\}}} 
\newcommand{\brk}[1]{\ensuremath{\left(#1\right)}} 
\newcommand{\op}[1]{\ensuremath{\texttt{#1}}}

\newcommand{\LS}[0]{\ensuremath{\mathcal{L}}\xspace} 
\newcommand{\PS}[0]{\ensuremath{\mathcal{K}}\xspace} 
\newcommand{\TS}[0]{\ensuremath{\mathcal{T}}\xspace} 

\newcommand{\BT}{\ensuremath{\mathcal{BT}}}
\newcommand{\VT}{\ensuremath{\mathcal{NT}}}
\newcommand{\ET}{\ensuremath{\mathcal{ET}}}

\newcommand{\elt}[1]{\ensuremath{\text{\cypherinline{#1}}}} 

\title{Schema Validation and Evolution for Graph Databases}

\date{}
\newcommand{\Objs}{\mathcal{O}}
\newcommand{\Keys}{\mathcal{K}}
\newcommand{\Nval}{\mathcal{V}}
\newcommand{\ie}{i.e.~}
\newcommand{\eg}{e.g.~}

\begin{document}


\title{Schema Validation and Evolution for Graph Databases}
\author{Angela Bonifati$^1$\footnote{Currently on leave at the University of 
Waterloo, Canada.}, Peter Furniss$^2$, Alastair Green$^2$, \\ 
Russ Harmer$^3$, Eugenia Oshurko$^3$, Hannes Voigt$^2$ \\ \\
$^1$ University of Lyon 1 \& CNRS LIRIS \\
$^2$ Neo4j \\
$^3$ CNRS \& ENS Lyon \& UCBL1}

\maketitle

\begin{abstract}
Despite the maturity of commercial graph databases, little consensus has been reached so far on the standardization of data definition languages (DDLs) for property graphs (PG). 
The discussion on the characteristics of PG schemas is ongoing in many standardization and community groups. 
Although some basic aspects of a schema are already present in Neo4j 3.5, like in most commercial graph databases, full support is missing allowing to constraint property graphs with more or less flexibility.

In this paper, we focus on two different perspectives from which a PG schema should be considered, as being descriptive or prescriptive, and we show how it would be possible to switch from one to another as the application under development gains more stability. 
Apart from proposing concise schema DDL inspired by Cypher syntax, we show how schema validation can be enforced through homomorphisms between PG schemas and PG instances; and how schema evolution can be described through the use of graph rewriting operations. 
Our prototypical implementation demonstrates feasibility and shows the need of offering high-level query primitives to accommodate flexible graph schema requirements as showcased in our work. 
\end{abstract}

\section{Introduction}


Property graph databases are modern data management systems that use graph structures, such as nodes, edges and properties, to encode semantically complex data. 
Graph database technology has made tremendous progress with many commercial products---such as Neo4j, Oracle PGX, SAP HANA Graph, Redis Graph, Cypher for Apache Spark and TigerGraph---and yet little consensus has been reached so far on the standardization of graph data querying and manipulation or of data definition languages (DDLs). 

The aim of ISO SC32/WG3 is to develop a new international standardized query language---called GQL\footnote{\url{https://www.gqlstandards.org/}}---for property graphs, with support from the activities of the wider community such as OpenCypher\footnote{\url{http://www.opencypher.org/}} and G-Core~\cite{AnglesABBFGLPPS18}.
Standardization of graph data querying and manipulation is therefore well under way.
However, schema specification for graphs along with a proposal for a graph data definition language have only recently started to be discussed within a standardization working group of ISO SC32/WG3 as well as in community working groups.
Indeed, there are only a few examples of property graph systems offering schema and DDL, e.g. Neo4j's Cypher for Apache Spark and TigerGraph.

Neo4j 3.5 already provides the means to express certain basic aspects of schemas. Specifically, the use of \emph{unique property} and \emph{property existence} constraints---or, more generally, of \emph{node keys}---on node and edge labels enables us to \emph{enforce} nodes (or edges) to have certain properties that moreover uniquely characterize that node (or edge). 
However, this does not allow users to express more advanced aspects of schemas such as specifying, for a given node or edge label, the collection of all possible associated properties; or constraining whether or not an edge may exist between nodes with certain labels.

The schemas that (property) graph database systems typically provide are {\em descriptive} in the sense that they only reflect the data: the schema can be changed simply by manipulating the data instance directly with no particular restrictions on any such manipulations.
The flexibility that this entails is generally perceived as a valuable characteristic, particularly in the earlier stages of application development, and especially in conjunction with the now ubiquitous agile software development method. 
A system that allows for the structure of graph elements to be manipulated and refactored freely, as the understanding and modelling of an application's universe evolves, greatly simplifies the development process in its early stages. 

As applications mature, however, a gradual shift in priorities occurs. 
As a concept becomes more stable, well-established, and central in our data model, we must treat it with increasing caution when considering further modifications.
The demand for restrictive schema manipulation policies further increases when an application goes into production since data becomes precious and misshaped data can have large financial consequences.
By this stage, traditional {\em prescriptive} schemas are the more appropriate choice.

Fully-specified schemas are indeed required by several Neo4j use cases, including insurance and pharmaceutical customers and GDPR\footnote{\url{https://eugdpr.org/}} compliance setups. Contrarily to relational settings, the schema requirements are oftentimes 
imposed by the evolving application rather than being encoded in the database at the very beginning of the build cycle. As such, schema evolution is not handled as in the relational setting by a dedicated team, but rather by the application team/business units.

However, descriptive and prescriptive schemas are only the two extremes of a spectrum of practical agility needs that application teams need to deal with.
Moreover, applications in productive use still continue to evolve and some schema changes are inevitable.
As an application and its schema grows, some parts of the schema mature faster than others giving rise to a different trade-off, between modification flexibility and demand for restriction, within a single schema.
For instance, the beta-available and under-development recommendation system added to an established shopping system requires more flexibility in schema than the shopping cart and order processing component.
Additionally, certain parts of an application may require greater flexibility independently of their state of maturity.
For instance, the product catalogue of the shopping system requires prescriptive properties, such as article number and prices, next to the perpetually changing set of descriptive properties that its products exhibit.

Modern database systems should support all these scenarios and allow users to change their flexibility/restriction requirements.
Traditional database schema approaches have been shaped by requirements originating from applications and application development methods that were been state-of-the-art many decades ago.
Today, the situation is different and the number of attempts~\cite{Beckmann2006,Moon2009,Curino2013,Voigt2014,Herrmann2017b,Herrmann2018a} to work around the restrictions of traditional schema approaches encoded in many database systems provide evidence of this.
As standardization of schema and DDL for PG database systems is just starting, we have a golden opportunity to consider these modern requirements on schema agility capabilities from the very beginning.

With this paper, we propose an PG schema approach that aims to accommodate modern schema requirements for property graph databases and offers support for prescriptive as well as descriptive schema in a very flexible fashion.
We make the following specific contributions:
\begin{itemize}
	\item a schema model specifying labels and (mandatory) properties for nodes and edges with mixing composition, guaranteed to be backward compatible with the flexible use of labels in today's PG databases while still facilitating strict typing of every graph element (Section \ref{S:DDL});
	\item a concise schema DDL with visually intuitive ASCII-art syntax inspired by Cypher (Section \ref{S:DDL});
	\item a mathematical framework for schema validation allowing us to construct both instances and schemas as property graphs and to enforce schema validation through the existence of a homomorphism from instance to schema (Section \ref{S:valid}); 
	\item mathematically specified graph rewriting rules \cite{corradini2006sesqui} and their application to update instances and/or schemas and propagate these changes from schema to instance (or vice versa) while keeping the instance and schema consistent at all times (Section \ref{S:PG-rewriting});
	\item a discussion of the requirements for graph refactoring and schema evolution based on the use of graph rewriting to express such operations mathematically (Sections \ref{S:SML}, \ref{S:evol} and \ref{S:discussion});
	\item a prototypical implementation demonstrating feasibility and showing the need of offering high-level primitives for schema validation and evolution in a PG query language; such prototypes builds on a Python library, called \texttt{ReGraph}, that allows rewriting and propagation by means of Cypher queries using Neo4j as backend (Section \ref{S:implem}). 
\end{itemize}
\vspace{-\topsep}
\section{PG Schema Language}\label{S:DDL}
We introduce in this section an OpenCypher-based schema DDL for Property Graphs (PG). 
Such a DDL is the outcome of extensive discussions at Neo4j about the graph DDL requirements and the possible extension of OpenCypher. 
Although informing and feeding the ongoing standardization process, our DLL must not be intended as a standard proposal since its main purpose is to substantiate the 
algorithmic contributions presented in the remainder of the paper.
The basic components of a schema definition assume a finite set of {\em labels}~\LS, a set of {\em property keys}~\PS and a finite set of data types~\TS.

\vspace{3pt}

\noindent{\textbf{Property graph type.}}
A property graph type is a triple \brk{\BT,\VT,\ET} where \BT{} is a set of element types, \VT{} is a set of node types and \ET{} is a set of edge types.
A property graph type provides the schema for a PG.
Multiple PGs can share a property graph type to the effect that they will have the same schema.

\vspace{3pt}

\noindent {\textbf{Property type.}}
A property type is a pair \brk{k,t}, where $k \in \PS$ is the property key and $t \in \TS$ is its data type. 
For instance, ``\cypherinline{content: STRING}'' declares the property type $\brk{\elt{content},\elt{STRING}}$.

\vspace{3pt}

\noindent {\textbf{Element type.}} 
An element type $b\in\BT$ is a 4-tuple $\brk{l,P,M,E}$, where $l \in \LS$ is a label, $P$ is a set of property types, $M \subseteq P$ is a subset of mandatory property types and $E\subseteq\BT$ is the set of element types that $b$ extends.

Hence, ``\cypherinline{Message \{content: STRING?,} \cypherinline{length: INTEGER\}}'' is a declaration of the element type $m=(\elt{Message},\set{pt_1,pt_2},$ $\set{pt_2},\emptyset)$, where $pt_1=(\elt{content},$ \\ $\elt{STRING})$ and $p_2=(\elt{length}, $ $\elt{INTEGER})$;
while ``\cypherinline{Post :: Message \{language: } \cypherinline{STRING?\}}'' declares the element type $p=(\elt{Post},\set{pt_3=\brk{\elt{language},\elt{STRING}}},$ $\emptyset,\set{m})$.

An element type is allowed to extend multiple other element types, but must not extend itself either directly or indirectly. 
All element types of a property graph type must be disambiguated by their label.
Where clear from context, we use the label to denote the corresponding element type.

\vspace{3pt}

\noindent {\textbf{Exposed (mandatory) property types and labels.}}
The set of exposed property types of an element type $b=(l, P, M, $ $E)$ is defined as $\op{prop}(b) := P \cup \bigcup_{c \in E}\op{prop}(c)$, \ie all the property types that $b$ possesses, either directly or through inheritance.
Similarly, we define $\op{mand}(b)$ to be the set of exposed mandatory property types of $b$ and $\op{labels}(b)$ to be the set of exposed labels of $b$. For instance, for element type $p$ from above we have $\op{prop}(p) = \set{pt_1,pt_2,pt_3}$, $\op{mand}(p) = \set{pt_2}$, and $\op{labels}(p) = \set{\elt{Post},\elt{Message}}$.

For an element type $b$ to be valid, $\op{prop}(b)$ must not have two or more property types with the same property key, \ie all properties types of a element type are disambiguated by their property key.
Where clear from context, we will use the property key to denote the corresponding property type. For instance, for the element type $p$ above, we have $\op{prop}(p) = \set{\elt{content},\elt{length},\elt{language}}$, $\op{mand}(p) = \set{\elt{length}}$ and $\op{labels}(p) = \{\elt{Post},$ \\
$\elt{Message}\}$.
Note that $\op{labels}(b)$ is unambiguous for all element types $b$ of a property graph type.

\vspace{3pt}

\noindent {\textbf{Node type.}}
A node type $nt\in\ET$ is a 1-tuple \brk{b}, where $b\in\BT$ is an element type.
For instance, ``\cypherinline{(Post)}'' declares the node type $p'=\brk{\elt{Post}}$. 
For a node type $nt=\brk{b}$, we define $\op{prop}(nt)=\op{prop}(b)$, $\op{mand}(nt)=\op{mand}(b)$, and $\op{labels}(nt)=\op{labels}(b)$.

\vspace{3pt}

\noindent {\textbf{Edge type.}}
An edge type $et\in\ET$ is a triple \brk{s,b,t}, where $s$, $b$, and $t$ are element types.
For instance, the edge type $(\elt{Comment},\elt{REPLY_OF},\elt{Message})$ can be declared with ``\cypherinline{(Comment)-[REPLY_OF]->(Message)}''.
Exposed (mandatory) property and label sets are defined analogously to node types based on $b$.
Note that $s$ and $t$ need not be node types. This allows a single edge type to be inherited by multiple node types.

\vspace{3pt}

\noindent {\textbf{Example.}}
The following snippet of the OpenCypher PG schema DDL creates a property graph type that captures an excerpt of the LDBC SNB~\cite{ErlingALCGPPB15} schema \footnote{The complete PG schema encoding of LDBC SNB is illustrated in the Appendix}. 
\begin{lstlisting}
CREATE GRAPH TYPE snb (
  // element types
  Person {
    firstName : STRING, lastName : STRING
  },
  Message { 
    creationDate : TIMESTAMP, browserUsed : STRING
  },
  Comment <: Message {},
  Post <: Message { 
  	imageFile : STRING?
  },
  REPLY_OF {},
  // node types
  (Person), (Post), (Comment),
  // edge types
  (Person)-[KNOWS]->(Person),
  (Person)-[LIKES]->(Message),
  (Message)-[HAS_CREATOR]->(Person),
  (Comment)-[REPLY_OF]->(Message)
)
\end{lstlisting}


%
%
%


\section{Schema Validation}\label{S:valid}

In this section, we provide a mathematical formalization of our notion of schema that, in particular, allows us to interpret a DDL specification as a PG. Apart from providing an intuitive visualization of the property graph type, this allows us to use the formalism for rewriting schemas in Sections \ref{S:PG-rewriting} and \ref{S:evol}.

Schema validation according to which an instance graph respects the schema can then be viewed as a homomorphism, \ie a structure-preserving function, from the instance to the schema. We present the mathematical definitions of schemas and instances as property graphs in Section \ref{S:pp} and then discuss the application of homomorphisms to the schema validation problem in Section \ref{S:schema-validation}.

\subsection{Schemas and instances as property graphs}
\label{S:pp}

We fix countable sets $\Objs$, $\Keys$ and $\Nval$ of \emph{objects}, \emph{keys} and \emph{values} respectively.
For the purposes of this paper, we assume that $\Nval$ contains (at least) basic types of integers, booleans, strings and dates.

A \emph{property graph} is defined to be a tuple $(N,E,\eta,P,\nu,M)$ where $N$ and $E$ are disjoint, finite subsets of $\Objs$ called \emph{nodes} and \emph{edges}; $\eta : E \rightarrow N \times N$ is a function assigning a source and target node to each edge; $P \subseteq (N \cup E) \times \Keys$ is a finite set of \emph{properties}; $\nu \subseteq P \times \Nval$ is a finite relation, assigning \emph{sets of values} to properties; and $M \subseteq P$ is a set of \emph{mandatory} properties.
The requirement that $\nu$ be finite means that each node and each edge has finitely many properties, each of which has a finite set of associated values.

\begin{figure}[h]
	\centering
	\scalebox{0.9}{
\begin{tikzpicture}[person/.style={draw, circle, fill=red!40!white}, post/.style={draw, circle, fill=blue!40!white}, comment/.style={draw, circle, fill=yellow!40!white}, attributes/.style={draw},attribute edge/.style={-, black!40!white}, graph edge/.style={-latex}, neutral/.style={draw, circle, fill=black!20!white}]
	\linespread{0.5}
	
	\node[person, minimum width=1.4cm] (person) at (0, 0) {\scriptsize Person};
  	\node[draw, minimum width=1.5cm, align=left] (person_attrs) at (-2, 0) {\tiny firstName:\ \texttt{STRING}\\ \tiny lastName:\ \texttt{STRING}};	
		
	\node[post, minimum width=1.4cm] (post) at (3, 1.25) {\scriptsize Post};
	\node[draw, minimum width=1.5cm, align=left] (post_attrs) at (1, 2.1) {\tiny imageFile:\ \texttt{STRING?}\\ \tiny creationDate:\ \texttt{STRING}\\ \tiny browserUsed:\ \texttt{STRING}};	
	 
	\node[comment, minimum width=1.4cm] (comment) at (3, -1.25) {\scriptsize Comment};
	\node[draw, minimum width=1.3cm, align=left] (comment_attrs) at (1, -2) {\tiny creationDate:\ \texttt{STRING}\\ \tiny browserUsed:\ \texttt{STRING}};
	
	\draw[-, black!40!white] (person_attrs) edge (person);
	\draw[-, black!40!white] (post_attrs) edge (post);
	\draw[-, black!40!white] (comment_attrs) edge (comment);

	\draw[-latex] (person) edge[out=115,in=75,looseness=7] node[midway,fill=white, sloped, pos=0.2] {\tiny \texttt{KNOWS}}  (person);
	\draw[-latex] (post) edge[bend right=15] node[midway, sloped, fill=white] {\tiny \texttt{HAS\_CREATOR}} (person);
	\draw[-latex] (comment) edge[bend left=15] node[midway, sloped, fill=white] {\tiny \texttt{HAS\_CREATOR}} (person);
	\draw[-latex] (person) edge[bend right=15] node[midway, sloped, fill=white] {\tiny \texttt{LIKES}} (post);
	\draw[-latex] (person) edge[bend left=15] node[midway,sloped, fill=white] {\tiny \texttt{LIKES}} (comment);
	\draw[-latex] (comment) edge[bend right=35] node[midway, sloped, fill=white, inner sep=1pt] {\tiny \texttt{REPLY\_OF}} (post);
	\draw[-latex] (comment) edge[out=30,in=-15, looseness=7] node[midway, sloped, pos=0.16, fill=white, inner sep=1pt] {\tiny \texttt{REPLY\_OF}}  (comment);

\end{tikzpicture}}
	\caption{An extract from the SNB schema} 
	\label{F:SNB}
\end{figure}
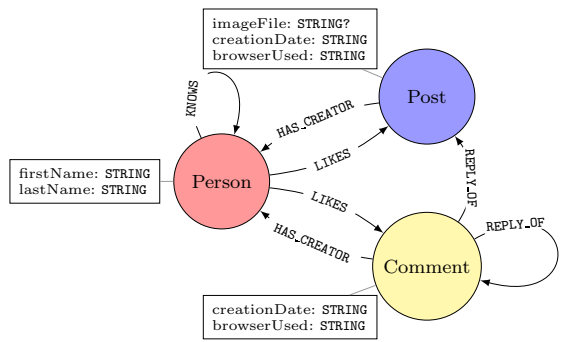

A schema \brk{\BT,\VT,\ET} specified in our DDL from Section \ref{S:DDL} can be interpreted as a property graph $S$ in the following way. The nodes $N_S$ are the node types \VT and we have an edge $e \in E_S$ from $n_1$ to $n_2$ in $E_S$ if, for some $l_1 \in \op{labels}(n_1)$ and $l_2 \in \op{labels}(n_2)$, there is an edge type $(n_1,e,n_2) \in \ET$. Note that a node type always gives rise to a single node of $S$ whereas an edge type may give rise to many edges in the schema graph; this is how inheritance in the DDL syntax is `expanded out' in the schema graph $S$ interpreting the property graph type. Each node and edge has the (mandatory) properties specified by its corresponding node or edge type.
As an example, the schema defined in Section \ref{S:DDL} and interpreted as a property graph is illustrated in Figure~\ref{F:SNB}.

\setlength{\textfloatsep}{10pt}

In this paper, we restrict our attention to \emph{simple} graphs, \ie $\eta$ is injective so we do not have parallel edges. This simplifies some of the technical details but it would be a straightforward matter to extend our results and implementation to the general case.

However, our definition of property graph has three differences from that found in \cite{2018Bonifati}: we have removed node and edge labels; we have added the notion of mandatory property; and we allow $\nu$ to be multi-valued rather than just a single-valued partial function.

In our mathematical framework, a graph schema and a graph instance are both represented as property graphs. The designation of one as the schema $S$ and the other as an instance $G$ is determined by the fact that we can map the latter to the former in a way that respects the graph structure. This is why we have removed labels from the definition of property graph: we can now think of the label( set)s of $G$ as \emph{being} the nodes/edges of $S$.

However, this notion of schema validation only allows us to express \emph{optional} properties; as such, we have added the notion of mandatory property to our notion of property graph, as discussed in Section \ref{S:schema-validation}, to be able to enforce the presence of properties in $G$.

The third difference arises due to the unavoidable fact that an update, or rewrite, of a property graph may cause a property to become associated with more than one value, as discussed in Section \ref{S:PG-rewriting}, typically due to the merging of nodes.

\subsection{Schema validation via graph homomorphisms}\label{S:schema-validation}

Let $G$ and $S$ be property graphs where $N_G \cup E_G$ and $N_S \cup E_S$ are disjoint. A \emph{homomorphism} $h : G \rightarrow S$ is a function $h_{\mathcal{N}} : N_G \rightarrow N_S$ and a function $h_{\mathcal{E}} : E_G \rightarrow E_S$, mapping nodes and edges of $G$ to nodes and edges of $S$, such that $\eta_S \circ h_{\mathcal{E}} = (h_{\mathcal{N}} \times h_{\mathcal{N}}) \circ \eta_G$. We write $h := h_{\mathcal{N}} \cup h_{\mathcal{E}}$. We further require that (i) if $(x,k) \in P_G$ then $(h(x),k) \in P_S$; (ii) if $((x,k),v) \in \nu_G$ then $((h(x),k),v) \in \nu_S$; and (iii) if $(h(x),k) \in M_S$ then $(x,k) \in M_G$.

In words, each edge of $G$ with source and target nodes $n_1$ and $n_2$ is mapped to an edge of $S$ with source and target nodes $h_{\mathcal{N}}(n_1)$ and $h_{\mathcal{N}}(n_2)$. We further require that (i) all properties of $G$ are instances of properties of $S$; (ii) each property in $G$ is associated with a subset of the values associated with its corresponding property in $S$; and (iii) any instance of a mandatory property of $S$ must be mandatory in $G$.

In the case of simple graphs, we do not need to specify the second function $h_{\mathcal{E}}$; it is enough to ask that, for all pairs $(n_1,n_2)$ of nodes of $G$, if there is an edge from $n_1$ to $n_2$ in $G$ then there must \emph{exist} a (necessarily unique) edge from $h_{\mathcal{N}}(n_1)$ to $h_{\mathcal{N}}(n_2)$.

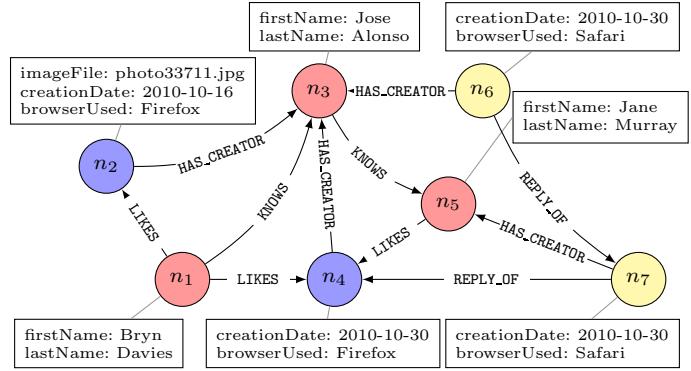
\begin{figure}[h]
	\centering
	\begin{tikzpicture}[person/.style={draw, circle, fill=red!40!white}, post/.style={draw, circle, fill=blue!40!white}, comment/.style={draw, circle, fill=yellow!40!white}, attributes/.style={draw},attribute edge/.style={-, black!40!white}, graph edge/.style={-latex}, neutral/.style={draw, circle, fill=black!20!white}]
	\linespread{0.5}
	
	\node[person] (n1) at (0, 0) {\scriptsize $n_1$};
	\node[draw, minimum width=1.5cm, align=left] (n1_attrs) at (-1.1, -0.85) {\tiny firstName:\ Bryn\\ \tiny lastName:\ Davies};	
	
	\node[post] (n2) at (-1, 1.5) {\scriptsize $n_2$};
	\node[draw, minimum width=1.5cm, align=left] (n2_attrs) at (-0.65, 2.5) {\tiny imageFile:\ photo33711.jpg \\ \tiny creationDate:\ 2010-10-16\\ \tiny browserUsed:\ Firefox};	
	
	\node[person] (n3) at (1.8, 2.5) {\scriptsize $n_3$};
	\node[draw, minimum width=1.5cm, align=left] (n3_attrs) at (2, 3.35) {\tiny firstName:\ Jose\\ \tiny lastName:\ Alonso};	
	
	\node[post] (n4) at (2, 0) {\scriptsize $n_4$};
	\node[draw, minimum width=1.5cm, align=left] (n4_attrs) at (1.85, -0.85) {\tiny creationDate:\ 2010-10-30\\ \tiny browserUsed:\ Firefox};		
	
	\node[person] (n5) at (3.5, 1) {\scriptsize $n_5$};
	\node[draw, minimum width=1.5cm, align=left] (n5_attrs) at (5.5, 2.15) {\tiny firstName:\ Jane\\ \tiny lastName:\ Murray};		
	
	\node[comment] (n6) at (3.95, 2.5) {\scriptsize $n_6$};
	\node[draw, minimum width=1.5cm, align=left] (n6_attrs) at (5, 3.35) {\tiny creationDate:\ 2010-10-30\\ \tiny browserUsed:\ Safari};	
	
	\node[comment] (n7) at (6, 0) {\scriptsize $n_7$};
	\node[draw, minimum width=1.5cm, align=left] (n7_attrs) at (5, -0.85) {\tiny creationDate:\ 2010-10-30\\ \tiny browserUsed:\ Safari};		
	
	\draw[-, black!40!white] (n1_attrs) edge (n1);
	\draw[-, black!40!white] (n2_attrs) edge (n2);
	\draw[-, black!40!white] (n3_attrs) edge (n3);
	\draw[-, black!40!white] (n4_attrs) edge (n4);
	\draw[-, black!40!white] (n6_attrs) edge (n6);	
	\draw[-, black!40!white] (n5_attrs.west) edge (n5);	
	\draw[-, black!40!white] (n7_attrs) edge (n7);	
	
	\draw[-latex] (n1) edge node[midway, sloped, fill=white] {\tiny \texttt{LIKES}}  (n2);
	\draw[-latex] (n1) edge[bend right=20] node[midway, sloped, fill=white] {\tiny \texttt{KNOWS}}  (n3);
	\draw[-latex] (n3) edge[bend right=20] node[midway, sloped, fill=white] {\tiny \texttt{KNOWS}}  (n5);
	\draw[-latex] (n1) edge  node[midway, sloped, fill=white] {\tiny \texttt{LIKES}} (n4);
	\draw[-latex] (n2) edge[bend right=20]  node[inner sep=0.5pt, midway, sloped, fill=white] {\tiny \texttt{HAS\_CREATOR}}  (n3);
	\draw[-latex] (n4) edge node[inner sep=0.5pt, midway, sloped, fill=white] {\tiny \texttt{HAS\_CREATOR}}   (n3);
	\draw[-latex] (n6) edge  node[inner sep=0.5pt, midway, sloped, fill=white] {\tiny \texttt{HAS\_CREATOR}} (n3);
	\draw[-latex] (n5) edge  node[midway, sloped, fill=white] {\tiny \texttt{LIKES}} (n4);
	\draw[-latex] (n6) edge[bend right=15] node[midway, sloped, fill=white] {\tiny \texttt{REPLY\_OF}}  (n7);
	\draw[-latex] (n7) edge  node[inner sep=0.5pt, midway, sloped, fill=white] {\tiny \texttt{HAS\_CREATOR}}  (n5);
	\draw[-latex] (n7) edge node[midway, sloped, fill=white] {\tiny \texttt{REPLY\_OF}}  (n4);
\end{tikzpicture}
	\caption{A valid instance of the SNB schema extract}
	\label{F:SNB-instance}
\end{figure}

We can view a homomorphism $h : G \rightarrow S$ as a formalization of the notion \emph{schema validation}, \ie that $G$ respects the `schema' $S$: each node/edge $x$ of $G$ is an instance of the schema node/edge $h(x)$; edges in $S$ constrain which edges can exist in $G$; and properties that are mandatory in the schema $S$ are mandatory (so must occur) in $G$. In the example instance $G$ of Figure \ref{F:SNB-instance}, we have used colours to encode the homomorphism $h$, \ie all yellow nodes are \cypherinline{Comment}s, etc. In the DDL of Section \ref{S:DDL}, the fact that all element types are disambiguated by their label would also allow us to determine $h$ provided we include these labels in the instance $G$.

\vspace{3pt}

\noindent {\textbf{The \texttt{ReGraph library}.}}
The Python library \texttt{ReGraph}\footnote{\url{https://github.com/Kappa-Dev/ReGraph}} provides a prototypical implementation of the presented system. It enables us to construct property graphs and structure them into hierarchies (DAGs) of graphs via homomorphisms. Although \texttt{ReGraph} can handle arbitrary hierarchies of graphs, in this paper we limit our use of the library to the special case of two graphs connected by a single homomorphism, \ie $h : G \rightarrow S$. This is sufficient for our purposes since, as shown in Section \ref{S:valid}, we can express and enforce schema validation with a hierarchy of precisely this kind.

\section{Schema manipulation}\label{S:SML}

In order to support schema agility, as is needed by applications today, a graph database system should provide schema manipulation operations (SMOs).
The following schema modifications are the minimum a system should possess in order to provide good coverage of the modifications required in practice and theory~\cite{Curino2008,Herrmann2015a}.
\begin{description}
\item[Create.]
The user can add new element types, node types and edge types to the schema.
\item[Drop.]
The user can remove element types, node types and edge types from the schema.
\item[Rename.]
The user can rename the label of an element type or the property key of a property type.
\item[Change.]
The user can change element types by adding, removing or changing property types.
\item[Partitioning/Split.]
The user can split or partition a node or edge type into more fine-grained node or edge types.
This is a common operation when a schema grows and gets further normalized so as to separate concepts.
Partitioning can happen horizontally as well as vertically.
In a (horizontal) partitioning, the user requires the distinction of \cypherinline{Message} into \cypherinline{Post} and \cypherinline{Comment}.
In a (vertical) split, the user separates \cypherinline{Message} into \cypherinline{MessageHeader} and \cypherinline{MessageBody}.
\item[Union/Join.]
The user can join or union node and edge types into more coarse-grained node and edge types.
This is needed when a schema shrinks or conceptual distinctions are generalized.
In a (horizontal) union, the user gives up distinguishing \cypherinline{Post} and \cypherinline{Comment} to consider simply \cypherinline{Message}.
In a (vertical) join, the user gives up distinguishing \cypherinline{MessageHeader} and \cypherinline{MessageBody} to consider simply \cypherinline{Message}.
\end{description}

Depending on the current state of maturity of a schema, users may want perform such operations {\em from schema to data} or {\em from data to schema}.
From schema to data---the traditional prescriptive schema evolution---is desirable for mature schemas in productive systems.
The user specifies how the schema is to change and the system propagates these changes to the data.
From data to schema---the descriptive schema manipulation---is desirable for more agile scenarios where the schema simply follows the data.
In this case, the user manipulates (parts of) the data and the system propagates these change to the schema.
Additionally, users may want to restrict mature parts of the schema from being manipulated descriptively, \eg by marking individual node and edge types as {\em final}.

This two-way propagation, from schema to data and from data to schema, is the challenging part for a system.
In particular, schema manipulations such as Partitioning/Split and Union/Join imply non-trivial propagations.
In the next section, we present the mathematical groundwork for such propagations.

\section{Property graph rewriting}\label{S:PG-rewriting}

In this section, we introduce (sesqui-pushout) graph rewriting rules \cite{corradini2006sesqui} which are our basic ingredient for performing schema evolution. In our mathematical framework, they allow us to modify property graphs. Graph rewriting for PG schema evolution has been inspired by its use
in graph-based knowledge representation and update in biological networks \cite{basso2016knowledge,harmer2017rule,harmer2017bio}. 

We first introduce these rules and explain the semantics of the graph rewrites that they perform. Then, in Section \ref{S:PG-propagation}, we focus on the application of these rules to schema graphs and instance graphs by propagating the corresponding operations from schemas to instances and vice versa.

\subsection{Rewriting rules}

A \emph{rewriting rule} is defined by three property graphs---$L$, $P$ and $R$---and two homomorphisms $\ell : P \rightarrow L$ and $r : P \rightarrow R$. The graph $L$ is called the \emph{left-hand side} (LHS) of the rule; a \emph{matching} of the rule into a graph $G$ is specified by an \emph{injective} homomorphism $m : L \rightarrowtail G$ that preserves mandatory properties: if $(x,k) \in M_G$ then $(h(x),k) \in M_S$. The graph $P$ is called the \emph{preserved region} and $R$ is the \emph{right-hand side} (RHS) of the rule.

The effect of rewriting $G$ through the matching $m$ can be specified abstractly due to the existence of certain operations on property graphs: a generalized set intersection, called \emph{pullback}; a generalized set union, called \emph{pushout}; and a generalized set difference, called \emph{pullback complement}. However, we can give an equivalent but more concrete definition in terms of elementary transformations of $G$.

A rule is \emph{restrictive} if $r : P \rightarrow P$ is the identity function. We can `read off' statically, from such a rule, a collection of elementary \emph{deletions} and \emph{clones} as follows:
\begin{itemize}
\item
a node, edge or property that occurs in $L$ but is not in the image of $\ell$ should be \emph{deleted};
\item
a single node of $L$ that is the image of $n$ nodes in $P$ through $\ell$ should be \emph{cloned} $n-1$ times.
\end{itemize}

\setlength{\columnsep}{10pt}
\begin{wrapfigure}[5]{r}{2cm}
\vspace{-10pt}
{\small
\begin{tikzcd}
	L \arrow[d,rightarrowtail,"m"']  & P \arrow[l,"\ell"'] \arrow[d, rightarrowtail, blue, "m^-"] \\
	G & {\color{blue} G^-} 		\arrow[l,blue,"\ell^-"]
\end{tikzcd}
}
\end{wrapfigure}
A restrictive rewrite of $G$ is specified by the pullback complement (in blue) $P \rightarrowtail G^- \rightarrow G$ of $\ell$ and $m$, where $G^-$ is the result of applying the elementary transformations of the rule to $G$.
Concretely, failures of surjectivity of $\ell$ give rise to deletions while cloning arises from failures of injectivity.

A rule is \emph{expansive} if $\ell : P \rightarrow P$ is the identity function. Such a rule only performs \emph{additions} and \emph{merges} which we can `read off' statically as follows:
\begin{itemize}
\item
a node, edge or property that occurs in $R$ but is not in the image of $r$ should be \emph{added};
\item
$n$ nodes in $P$ that all map through $r$ to the same node of $R$ should be \emph{merged} into a single node.
\end{itemize}

\setlength{\columnsep}{10pt}
\begin{wrapfigure}[5]{r}{2.5cm}
{\small
	\begin{tikzcd}
		P \arrow[d,rightarrowtail,"m^-"'] \arrow[r,"r"] & R \arrow[d,rightarrowtail,blue,"m^+"] \\
		G^- \arrow[r,blue,"r^+"'] & {\color{blue} G^+}
	\end{tikzcd}
}
\end{wrapfigure}
An expansive rewrite of $G^-$ is specified by the pushout (in blue) $R \rightarrowtail G^+ \leftarrow G^-$ of $m^-$ and $r$, where $G^+$ is the result of applying the elementary transformations of the rule to $G^-$.
Concretely, failures of surjectivity in  $r$ give rise to addition while failures of injectivity give rise to merging.

The overall effect of a rule is determined by performing these two phases of rewriting consecutively: first the restrictive phase; then the expansive phase.

If the graph $G$ that we wish to rewrite respects the schema $S$, \ie we have a homomorphism $h : G \rightarrow S$, then the resulting $G^+$ will still respect $S$ provided that the rule respects $S$ in the following sense: 

\begin{wrapfigure}[5]{r}{2cm}
\vspace{-10pt}
{\small
	\begin{tikzcd}
		\arrow[d,"\ell"'] P \arrow[r,"r"] & R \arrow[d,"h_R"] \\
		L \arrow[r,"h_L"'] & S
	\end{tikzcd}
}
\end{wrapfigure}
\noindent we have homomorphisms $h_L : L \rightarrow S$ and from $h_R : R \rightarrow S$ such that $h_L \circ \ell = h_R \circ r$. In words, $L$ and $R$ respect $S$ individually and in such a way that they agree on their overlap $P$.

\vspace{3pt}

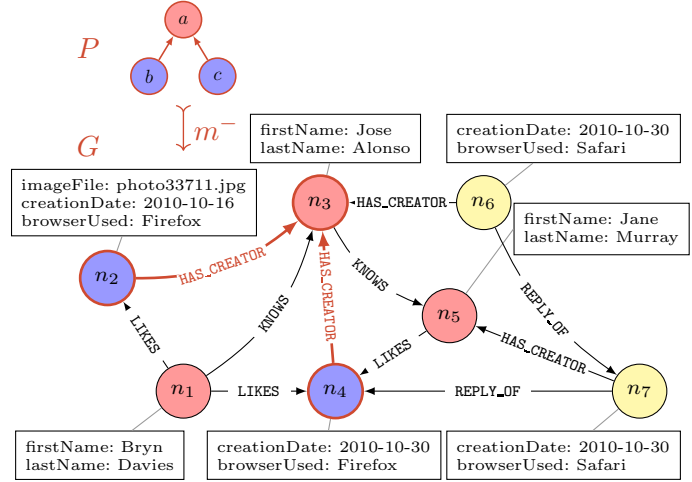
\begin{figure}[h]
	\centering
	\begin{tikzpicture}[person/.style={draw, circle, fill=red!40!white}, post/.style={draw, circle, fill=blue!40!white}, comment/.style={draw, circle, fill=yellow!40!white}, attributes/.style={draw},attribute edge/.style={-, black!40!white}, graph edge/.style={-latex}, neutral/.style={draw, circle, fill=black!20!white}]
	\linespread{0.5}
	
	\node (p_label) at (-1.25, 4.55) {\color{Orange} $P$};
	\node[scale=0.6] (p) at (0, 4.55) {
		\begin{tikzpicture}
	
			\node[person, draw=Orange, line width=1pt, minimum width=0.8cm] (pa) at (1.5, 6) {\large $a$};
			\node[post, draw=Orange, line width=1pt, minimum width=0.8cm] (pb) at (0.75, 4.75) {\large $b$};
			\node[post, draw=Orange, line width=1pt, minimum width=0.8cm] (pc) at (2.25, 4.75) {\large $c$};
	
			\draw[-latex, Orange, line width=1pt] (pb) edge (pa);
			\draw[-latex, Orange, line width=1pt] (pc) edge (pa);

		\end{tikzpicture}
	};

	\node (p_label) at (-1.25, 3.25) {\color{Orange} $G$};
	\node (phantom) at (0, 3) {};
	
	\node[person] (n1) at (0, 0) {\scriptsize $n_1$};
	\node[draw, minimum width=1.5cm, align=left] (n1_attrs) at (-1.1, -0.85) {\tiny firstName:\ Bryn\\ \tiny lastName:\ Davies};	
	
	\node[post, draw=Orange, line width=1pt] (n2) at (-1, 1.5) {\scriptsize $n_2$};
	\node[draw, minimum width=1.5cm, align=left] (n2_attrs) at (-0.65, 2.5) {\tiny imageFile:\ photo33711.jpg \\ \tiny creationDate:\ 2010-10-16\\ \tiny browserUsed:\ Firefox};	
	
	\node[person, draw=Orange, line width=1pt] (n3) at (1.8, 2.5) {\scriptsize $n_3$};
	\node[draw, minimum width=1.5cm, align=left] (n3_attrs) at (2, 3.35) {\tiny firstName:\ Jose\\ \tiny lastName:\ Alonso};	
	
	\node[post, draw=Orange, line width=1pt] (n4) at (2, 0) {\scriptsize $n_4$};
	\node[draw, minimum width=1.5cm, align=left] (n4_attrs) at (1.85, -0.85) {\tiny creationDate:\ 2010-10-30\\ \tiny browserUsed:\ Firefox};		
	
	\node[person] (n5) at (3.5, 1) {\scriptsize $n_5$};
	\node[draw, minimum width=1.5cm, align=left] (n5_attrs) at (5.5, 2.15) {\tiny firstName:\ Jane\\ \tiny lastName:\ Murray};		
	
	\node[comment] (n6) at (3.95, 2.5) {\scriptsize $n_6$};
	\node[draw, minimum width=1.5cm, align=left] (n6_attrs) at (5, 3.35) {\tiny creationDate:\ 2010-10-30\\ \tiny browserUsed:\ Safari};	
	
	\node[comment] (n7) at (6, 0) {\scriptsize $n_7$};
	\node[draw, minimum width=1.5cm, align=left] (n7_attrs) at (5, -0.85) {\tiny creationDate:\ 2010-10-30\\ \tiny browserUsed:\ Safari};		
	
	\draw[-, black!40!white] (n1_attrs) edge (n1);
	\draw[-, black!40!white] (n2_attrs) edge (n2);
	\draw[-, black!40!white] (n3_attrs) edge (n3);
	\draw[-, black!40!white] (n4_attrs) edge (n4);
	\draw[-, black!40!white] (n6_attrs) edge (n6);	
	\draw[-, black!40!white] (n5_attrs.west) edge (n5);	
	\draw[-, black!40!white] (n7_attrs) edge (n7);	
	
	\draw[-latex] (n1) edge node[midway, sloped, fill=white] {\tiny \texttt{LIKES}}  (n2);
	\draw[-latex] (n1) edge[bend right=20] node[midway, sloped, fill=white] {\tiny \texttt{KNOWS}}  (n3);
	\draw[-latex] (n3) edge[bend right=20] node[midway, sloped, fill=white] {\tiny \texttt{KNOWS}}  (n5);
	\draw[-latex] (n1) edge  node[midway, sloped, fill=white] {\tiny \texttt{LIKES}} (n4);
	\draw[-latex] (n2) edge[bend right=20, Orange, line width=1pt]  node[inner sep=0.5pt, midway, sloped, fill=white] {\tiny \texttt{HAS\_CREATOR}}  (n3);
	\draw[-latex, Orange, line width=1pt] (n4) edge node[inner sep=0.5pt, midway, sloped, fill=white] {\tiny \texttt{HAS\_CREATOR}}   (n3);
	\draw[-latex] (n6) edge  node[inner sep=0.5pt, midway, sloped, fill=white] {\tiny \texttt{HAS\_CREATOR}} (n3);
	\draw[-latex] (n5) edge  node[midway, sloped, fill=white] {\tiny \texttt{LIKES}} (n4);
	\draw[-latex] (n6) edge[bend right=15] node[midway, sloped, fill=white] {\tiny \texttt{REPLY\_OF}}  (n7);
	\draw[-latex] (n7) edge  node[inner sep=0.5pt, midway, sloped, fill=white] {\tiny \texttt{HAS\_CREATOR}}  (n5);
	\draw[-latex] (n7) edge node[midway, sloped, fill=white] {\tiny \texttt{REPLY\_OF}}  (n4);
	
	\draw[>->, Orange, line width=0.5pt, shorten <= 0.05cm, shorten >= 0.05cm] (p) edge node[right] {$m^-$} (phantom);
\end{tikzpicture}
	\caption{A matching of $P$ into the instance of Figure 2}
\label{F:rewrite1}
\end{figure}

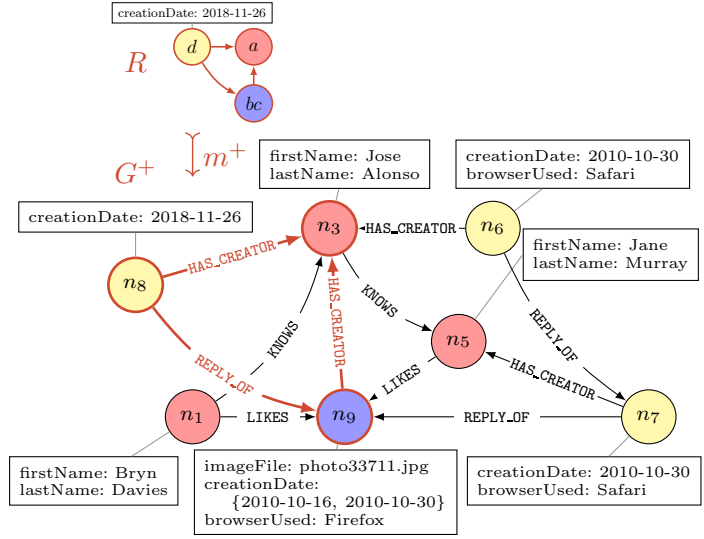
\begin{figure}[h]
	\centering
	\begin{tikzpicture}[person/.style={draw, circle, fill=red!40!white}, post/.style={draw, circle, fill=blue!40!white}, comment/.style={draw, circle, fill=yellow!40!white}, attributes/.style={draw},attribute edge/.style={-, black!40!white}, graph edge/.style={-latex}, neutral/.style={draw, circle, fill=black!20!white}]
	\linespread{0.5}

	\node (r_label) at (-0.75, 4.7) {\color{Orange} $R$};
	\node[scale=0.6] (p) at (0, 4.7) {
		\begin{tikzpicture}
	
			\node[person, draw=Orange, line width=1pt, minimum width=0.8cm] (ra) at (1.5, 6) {\large $a$};
	
			\node[comment, draw=Orange, line width=1pt, minimum width=0.8cm] (rd) at (0.15, 6) {\large $d$};
			\node[draw, minimum width=1.5cm, align=left, minimum width=1.5cm, align=left] (rd_attrs) at (0.15, 6.75) {\scriptsize creationDate:\ 2018-11-26};	
			
			\node[post, draw=Orange, line width=1pt, minimum width=0.8cm] (rbc) at (1.5, 4.75) {\large $bc$};
	
			\draw[-latex, Orange, line width=1pt] (rd) edge (ra);
			\draw[-latex, Orange, line width=1pt] (rbc) edge (ra);
			\draw[-latex, Orange, line width=1pt] (rd) edge[bend right=15] (rbc);
	
			\draw[-, black!40!white] (rd_attrs) edge (rd);
		\end{tikzpicture}
	};

	\node (r_label) at (-0.75, 3.25) {\color{Orange} $G^+$};
	\node (phantom) at (0, 3) {};
	
	\node[person] (n1) at (0, 0) {\scriptsize $n_1$};
	\node[draw, minimum width=1.5cm, align=left] (n1_attrs) at (-1.3, -0.85) {\tiny firstName:\ Bryn\\ \tiny lastName:\ Davies};	
	
	\node[comment, draw=Orange, line width=1pt] (n8) at (-0.75, 1.75) {\scriptsize $n_8$};
	\node[draw, minimum width=1.5cm, align=left] (n8_attrs) at (-0.75, 2.65) {\tiny creationDate:\ 2018-11-26};	
	
	\node[person, draw=Orange, line width=1pt] (n3) at (1.8, 2.5) {\scriptsize $n_3$};
	\node[draw, minimum width=1.5cm, align=left] (n3_attrs) at (2, 3.35) {\tiny firstName:\ Jose\\ \tiny lastName:\ Alonso};	
	
	\node[post, draw=Orange, line width=1pt] (n4) at (2, 0) {\scriptsize $n_9$};
	\node[draw, minimum width=1.5cm, align=left] (n4_attrs) at (1.75, -1) {\tiny imageFile:\ photo33711.jpg \\ \tiny creationDate:\ \\ 
	 \tiny\ \ \ \ \{2010-10-16, 2010-10-30\} \\
	 \tiny browserUsed:\ Firefox};		
	
	\node[person] (n5) at (3.5, 1) {\scriptsize $n_5$};
	\node[draw, minimum width=1.5cm, align=left] (n5_attrs) at (5.5, 2.15) {\tiny firstName:\ Jane\\ \tiny lastName:\ Murray};		
	
	\node[comment] (n6) at (3.95, 2.5) {\scriptsize $n_6$};
	\node[draw, minimum width=1.5cm, align=left] (n6_attrs) at (5, 3.35) {\tiny creationDate:\ 2010-10-30\\ \tiny browserUsed:\ Safari};	
	
	\node[comment] (n7) at (6, 0) {\scriptsize $n_7$};
	\node[draw, minimum width=1.5cm, align=left] (n7_attrs) at (5.15, -0.85) {\tiny creationDate:\ 2010-10-30\\ \tiny browserUsed:\ Safari};		
	
	\draw[-, black!40!white] (n1_attrs) edge (n1);
	\draw[-, black!40!white] (n8_attrs) edge (n8);
	\draw[-, black!40!white] (n3_attrs) edge (n3);
	\draw[-, black!40!white] (n4_attrs) edge (n4);
	\draw[-, black!40!white] (n6_attrs) edge (n6);	
	\draw[-, black!40!white] (n5_attrs.west) edge (n5);	
	\draw[-, black!40!white] (n7_attrs) edge (n7);	
	
	\draw[-latex] (n1) edge[bend right=20] node[midway, sloped, fill=white] {\tiny \texttt{KNOWS}}  (n3);
	\draw[-latex] (n3) edge[bend right=20] node[midway, sloped, fill=white] {\tiny \texttt{KNOWS}}  (n5);
	\draw[-latex] (n1) edge  node[midway, sloped, fill=white] {\tiny \texttt{LIKES}} (n4);
	\draw[-latex, Orange, line width=1pt] (n4) edge node[inner sep=0.5pt, midway, sloped, fill=white] {\tiny \texttt{HAS\_CREATOR}}   (n3);
	\draw[-latex] (n6) edge  node[inner sep=0.5pt, midway, sloped, fill=white] {\tiny \texttt{HAS\_CREATOR}} (n3);
	\draw[-latex] (n5) edge  node[midway, sloped, fill=white] {\tiny \texttt{LIKES}} (n4);
	\draw[-latex] (n6) edge[bend right=15] node[midway, sloped, fill=white] {\tiny \texttt{REPLY\_OF}}  (n7);
	\draw[-latex] (n7) edge  node[inner sep=0.5pt, midway, sloped, fill=white] {\tiny \texttt{HAS\_CREATOR}}  (n5);
	\draw[-latex] (n7) edge node[midway, sloped, fill=white] {\tiny \texttt{REPLY\_OF}}  (n4);
	
	\draw[-latex, Orange, line width=1pt] (n8) edge node[inner sep=0.5pt, midway, sloped, fill=white] {\tiny \texttt{HAS\_CREATOR}}  (n3);
   \draw[-latex, Orange, line width=1pt] (n8) edge[bend right=20]  node[midway, sloped, fill=white] {\tiny \texttt{REPLY\_OF}}  (n4);
	
	\draw[>->, Orange, line width=0.5pt, shorten <= 0.05cm, shorten >= 0.05cm] (p) edge node[right] {$m^+$} (phantom);
\end{tikzpicture}
	\caption{The matching of $R$ into the rewritten instance}
\label{F:rewrite2}
\end{figure}

\noindent {\textbf{Example.}}
Figure \ref{F:rewrite1} illustrates a graph $P$ and a matching of $P$ in the instance graph (see Figure \ref{F:SNB-instance}) where the nodes $a$, $b$ and $c$ of $P$ are mapped to $n_3$, $n_2$ and $n_4$ respectively. The graph $R$ in Figure \ref{F:rewrite2} is the RHS of an expansive rule $r : P \rightarrow R$ where $r$ is determined by the colour coding. Note that $r$ is not injective, as it maps $b$ and $c$ in $P$ to $bc$ in $R$; nor is it surjective as it maps nothing to $d$.

Figure \ref{F:rewrite2} also shows the matching of $R$ into the rewritten instance: $n_2$ and $n_4$ have been merged into $n_9$ and $n_8$ has been added along with its incident edges as specified in $R$.

\setlength{\textfloatsep}{0pt}

If the rule does \emph{not} respect $S$, we lose schema validation of $G^+$ with respect to $S$. In order to restore the required homomorphism, the rewrite of $G$ must be \emph{propagated} to $S$ (or vice versa).

\subsection{Propagation of rewriting}\label{S:PG-propagation}

We need to consider four kinds of rewrites: application of a \emph{restrictive} or an \emph{expansive} rule to $S$ or to $G$. If we rewrite $S$ to $S^+$ with an expansive rule, we immediately obtain a homomorphism $h^+ : G \rightarrow S^+$ by composition and no propagation to $G$ is required. The same applies to rewriting $G$ to $G^-$ with a restrictive rule.

The remaining cases are more involved. If we rewrite $S$ to $S^-$ with a \emph{restrictive} rule, a homomorphism from $G$ to $S^-$ may no longer exist since either: 
 \begin{itemize}
 \item[(i)] we have deleted an element, \ie a node or edge or property, of $S$ to which elements in $G$ were mapped; or 

 \item[(ii)] we have cloned a node in $S$ and so no longer know to which node we should map things in $G$. 
 \end{itemize}
However, we can determine a \emph{canonical} rewrite of $G$ to some $G^-$ that restores a homomorphism $h^- : G^- \rightarrow S^-$. The rewrite of $G$ is determined by propagating the elementary transformations of the rewrite of $S$: (i) a node of $G$ that is mapped by $h$ to a deleted node of $S$ is itself deleted; and (ii) a node of $G$ that is mapped by $h$ to a cloned node of $S$ is itself cloned (the same number of times).


Conversely, if we rewrite $G$ to $G^+$ with an \emph{expansive} rule, we may no longer have a homomorphism from $G^+$ to $S$ because 
\begin{itemize}
 \item[(i)] we have added an element to $G$ that we do not know how to map to $S$; or 
 
 \item[(ii)] we have merged nodes in $G$ that are mapped by $h$ to different nodes of $S$. 
\end{itemize}
However, we can again deduce a canonical rewrite of $S$ to some $S^+$ 
to restore a homomorphism $h^+ : G^+ \rightarrow S^+$.

\vspace{3pt}

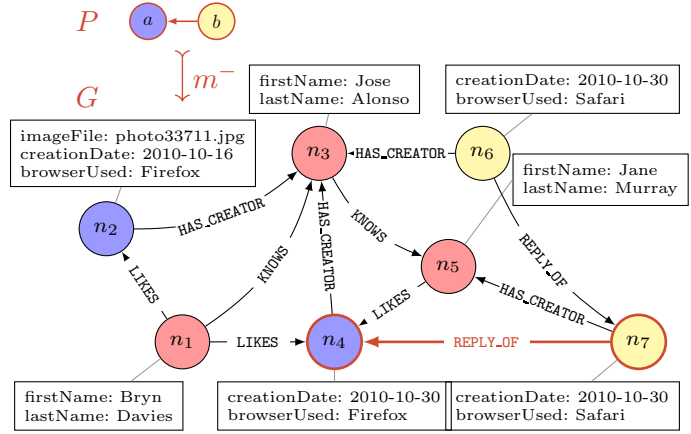
\begin{figure}[h]
	\centering
	\begin{tikzpicture}[person/.style={draw, circle, fill=red!40!white}, post/.style={draw, circle, fill=blue!40!white}, comment/.style={draw, circle, fill=yellow!40!white}, attributes/.style={draw},attribute edge/.style={-, black!40!white}, graph edge/.style={-latex}, neutral/.style={draw, circle, fill=black!20!white}]
	\linespread{0.5}
	
	\node (p_label) at (-1.25, 4.25) {\color{Orange} $P$};
	\node[scale=0.6] (p) at (0, 4.25) {
		\begin{tikzpicture}
			\node[post, draw=Orange, line width=1pt, minimum width=0.8cm] (pa) at (0, 4.75) {\large $a$};
	
			\node[comment, draw=Orange, line width=1pt, minimum width=0.8cm] (pb) at (1.5, 4.75) {\large $b$};	
			
			\draw[-latex, Orange, line width=1pt] (pb) edge (pa);
		\end{tikzpicture}
	};
	
	\node (p_label) at (-1.25, 3.25) {\color{Orange} $G$};
	\node (phantom) at (0, 3) {};
	
	\node[person] (n1) at (0, 0) {\scriptsize $n_1$};
	\node[draw, minimum width=1.5cm, align=left] (n1_attrs) at (-1.1, -0.85) {\tiny firstName:\ Bryn\\ \tiny lastName:\ Davies};	
	
	\node[post] (n2) at (-1, 1.5) {\scriptsize $n_2$};
	\node[draw, minimum width=1.5cm, align=left] (n2_attrs) at (-0.65, 2.5) {\tiny imageFile:\ photo33711.jpg \\ \tiny creationDate:\ 2010-10-16\\ \tiny browserUsed:\ Firefox};	
	
	\node[person] (n3) at (1.8, 2.5) {\scriptsize $n_3$};
	\node[draw, minimum width=1.5cm, align=left] (n3_attrs) at (2, 3.35) {\tiny firstName:\ Jose\\ \tiny lastName:\ Alonso};	
	
	\node[post, draw=Orange, line width=1pt] (n4) at (2, 0) {\scriptsize $n_4$};
	\node[draw, minimum width=1.5cm, align=left] (n4_attrs) at (2, -0.85) {\tiny creationDate:\ 2010-10-30\\ \tiny browserUsed:\ Firefox};		
	
	\node[person] (n5) at (3.5, 1) {\scriptsize $n_5$};
	\node[draw, minimum width=1.5cm, align=left] (n5_attrs) at (5.5, 2.15) {\tiny firstName:\ Jane\\ \tiny lastName:\ Murray};		
	
	\node[comment] (n6) at (3.95, 2.5) {\scriptsize $n_6$};
	\node[draw, minimum width=1.5cm, align=left] (n6_attrs) at (5, 3.35) {\tiny creationDate:\ 2010-10-30\\ \tiny browserUsed:\ Safari};	
	
	\node[comment, draw=Orange, line width=1pt] (n7) at (6, 0) {\scriptsize $n_7$};
	\node[draw, minimum width=1.5cm, align=left] (n7_attrs) at (5, -0.85) {\tiny creationDate:\ 2010-10-30\\ \tiny browserUsed:\ Safari};		
	
	\draw[-, black!40!white] (n1_attrs) edge (n1);
	\draw[-, black!40!white] (n2_attrs) edge (n2);
	\draw[-, black!40!white] (n3_attrs) edge (n3);
	\draw[-, black!40!white] (n4_attrs) edge (n4);
	\draw[-, black!40!white] (n6_attrs) edge (n6);	
	\draw[-, black!40!white] (n5_attrs.west) edge (n5);	
	\draw[-, black!40!white] (n7_attrs) edge (n7);	
	
	\draw[-latex] (n1) edge node[midway, sloped, fill=white] {\tiny \texttt{LIKES}}  (n2);
	\draw[-latex] (n1) edge[bend right=20] node[midway, sloped, fill=white] {\tiny \texttt{KNOWS}}  (n3);
	\draw[-latex] (n3) edge[bend right=20] node[midway, sloped, fill=white] {\tiny \texttt{KNOWS}}  (n5);
	\draw[-latex] (n1) edge  node[midway, sloped, fill=white] {\tiny \texttt{LIKES}} (n4);
	\draw[-latex] (n2) edge[bend right=20]  node[inner sep=0.5pt, midway, sloped, fill=white] {\tiny \texttt{HAS\_CREATOR}}  (n3);
	\draw[-latex] (n4) edge node[inner sep=0.5pt, midway, sloped, fill=white] {\tiny \texttt{HAS\_CREATOR}}   (n3);
	\draw[-latex] (n6) edge  node[inner sep=0.5pt, midway, sloped, fill=white] {\tiny \texttt{HAS\_CREATOR}} (n3);
	\draw[-latex] (n5) edge  node[midway, sloped, fill=white] {\tiny \texttt{LIKES}} (n4);
	\draw[-latex] (n6) edge[bend right=15] node[midway, sloped, fill=white] {\tiny \texttt{REPLY\_OF}}  (n7);
	\draw[-latex] (n7) edge  node[inner sep=0.5pt, midway, sloped, fill=white] {\tiny \texttt{HAS\_CREATOR}}  (n5);
	\draw[-latex, draw=Orange, line width=1pt] (n7) edge[Orange] node[midway, sloped, fill=white] {\tiny \texttt{REPLY\_OF}}  (n4);
	
	\draw[>->, Orange, line width=0.5pt, shorten <= 0.05cm, shorten >= 0.05cm] (p) edge node[right] {$m^-$} (phantom);
\end{tikzpicture}
	\caption{A matching of $P$ into the instance of Figure 2}
	\label{F:rewrite3}
\end{figure}

\begin{figure}[h]
	\centering
	\scalebox{0.85}{\begin{tikzpicture}[person/.style={draw, circle, fill=red!40!white}, message/.style={draw, circle, fill=green!80!blue!40!white}, post/.style={draw, circle, fill=blue!40!white}, comment/.style={draw, circle, fill=yellow!40!white}, attributes/.style={draw},attribute edge/.style={-, black!40!white}, graph edge/.style={-latex}, neutral/.style={draw, circle, fill=black!20!white}]
	\linespread{0.5}
	
	\node[person, minimum width=1.8cm] (person) at (-0.25, 0) {Person};
  	\node[draw, minimum width=1.5cm, align=left] (person_attrs) at (-2.5, 0) {\tiny firstName:\ \texttt{STRING}\\ \tiny lastName:\ \texttt{STRING}};	
		
	\node[message, minimum width=1.5cm] (message) at (3, 0) {Message};
	\node[draw, minimum width=1.5cm, align=left] (message_attrs) at (5.25, 0) {\tiny imageFile:\ \texttt{STRING?}\\ \tiny creationDate:\ \texttt{STRING}\\  \tiny browserUsed:\ \texttt{STRING}};	
	
	\draw[-, black!40!white] (person_attrs) edge (person);
	\draw[-, black!40!white] (message_attrs) edge (message);

	\draw[-latex] (person) edge[out=115,in=65, looseness=5] node[midway,fill=white, pos=0.16, sloped, inner sep=1pt] {\tiny \texttt{KNOWS}}  (person);
	\draw[-latex] (message) edge[bend right=15] node[midway,fill=white, inner sep=1pt] {\tiny \texttt{HAS\_CREATOR}} (person);
	\draw[-latex] (person) edge[bend right=15] node[midway,fill=white, inner sep=1pt] {\tiny \texttt{LIKES}} (message);
	\draw[-latex] (message) edge[out=115,in=65, looseness=5] node[midway,fill=white, pos=0.16, sloped, inner sep=1pt] {\tiny \texttt{REPLY\_OF}}  (message);

\end{tikzpicture}}
	\caption{The schema after propagation of rewriting}
	\label{F:new-schema}
\end{figure}
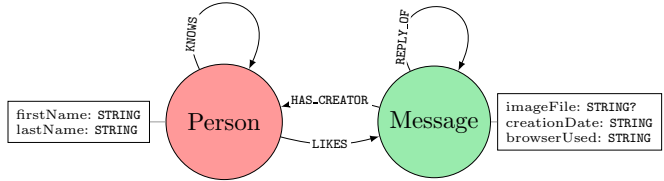

\noindent {\textbf{Example.}}
The propagation of a merge to $S$ may change the type of many nodes in $G$---including those not directly affected by the rewrite. In Figures \ref{F:rewrite3} and \ref{F:rewrite4}, the rule merges a post and a comment in $G$ but propagation to $S$ has the side-effect that \emph{all} posts and comments, not just the merged node $n_8$, now map to the \cypherinline{Message} node of the updated schema in Figure \ref{F:new-schema}.

\begin{figure}[h]
	\centering
	\begin{tikzpicture}[person/.style={draw, circle, fill=red!40!white}, message/.style={draw, circle, fill=green!80!blue!40!white}, post/.style={draw, circle, fill=blue!40!white}, comment/.style={draw, circle, fill=yellow!40!white}, attributes/.style={draw},attribute edge/.style={-, black!40!white}, graph edge/.style={-latex}, neutral/.style={draw, circle, fill=black!20!white}]
	\linespread{0.5}
	
	\node (r_label) at (-1, 4.25) {\color{Orange} $R$};
	\node[scale=0.6] (p) at (0, 4.25) {
		\begin{tikzpicture}
			\node[message, draw=Orange, line width=1pt, minimum width=0.8cm] (pa) at (0, 4.75) {\large $ab$};
	
			\draw[-latex, Orange, line width=1pt] (pa) edge[out=115, in=65, looseness=5] (pa);
		\end{tikzpicture}
	};
	
	\node (r_label) at (-1, 3.25) {\color{Orange} $G^+$};	
	\node (phantom) at (0, 3) {};
	
	\node[person] (n1) at (0, 0) {\scriptsize $n_1$};
	\node[draw, minimum width=1.5cm, align=left] (n1_attrs) at (-1.1, -0.85) {\tiny firstName:\ Bryn\\ \tiny lastName:\ Davies};	
	
	\node[message] (n2) at (-1, 1.5) {\scriptsize $n_2$};
	\node[draw, minimum width=1.5cm, align=left] (n2_attrs) at (-0.65, 2.5) {\tiny imageFile:\ photo33711.jpg \\ \tiny creationDate:\ 2010-10-16\\ \tiny browserUsed:\ Firefox};	
	
	\node[person] (n3) at (1.8, 2.5) {\scriptsize $n_3$};
	\node[draw, minimum width=1.5cm, align=left] (n3_attrs) at (2, 3.35) {\tiny firstName:\ Jose\\ \tiny lastName:\ Alonso};	
	
	\node[message, draw=Orange, line width=1pt] (n8) at (2, 0) {\scriptsize $n_8$};
	\node[draw, minimum width=1.5cm, align=left] (n8_attrs) at (2, -0.85) {\tiny creationDate:\ 2010-10-30\\ \tiny browserUsed:\ \{Firefox, Safari \}};		
	
	\node[person] (n5) at (4.5, 1) {\scriptsize $n_5$};
	\node[draw, minimum width=1.5cm, align=left] (n5_attrs) at (5.5, 2.15) {\tiny firstName:\ Jane\\ \tiny lastName:\ Murray};		
	
	\node[message] (n6) at (3.95, 2.5) {\scriptsize $n_6$};
	\node[draw, minimum width=1.5cm, align=left] (n6_attrs) at (5, 3.35) {\tiny creationDate:\ 2010-10-30\\ \tiny browserUsed:\ Safari};	
	
	
	\draw[-, black!40!white] (n1_attrs) edge (n1);
	\draw[-, black!40!white] (n2_attrs) edge (n2);
	\draw[-, black!40!white] (n3_attrs) edge (n3);
	\draw[-, black!40!white] (n8_attrs) edge (n8);
	\draw[-, black!40!white] (n6_attrs) edge (n6);	
	\draw[-, black!40!white] (n5_attrs) edge (n5);	
	
	\draw[-latex] (n1) edge node[midway, sloped, fill=white] {\tiny \texttt{LIKES}}  (n2);
	\draw[-latex] (n1) edge[bend right=20] node[midway, sloped, fill=white] {\tiny \texttt{KNOWS}}  (n3);
	\draw[-latex] (n3) edge[bend right=20] node[midway, sloped, fill=white] {\tiny \texttt{KNOWS}}  (n5);
	\draw[-latex] (n1) edge  node[midway, sloped, fill=white] {\tiny \texttt{LIKES}} (n8);
	\draw[-latex] (n2) edge[bend right=20]  node[inner sep=0.5pt, midway, sloped, fill=white] {\tiny \texttt{HAS\_CREATOR}}  (n3);
	\draw[-latex] (n8) edge node[inner sep=0.5pt, midway, sloped, fill=white] {\tiny \texttt{HAS\_CREATOR}}   (n3);
	\draw[-latex] (n6) edge  node[inner sep=0.5pt, midway, sloped, fill=white] {\tiny \texttt{HAS\_CREATOR}} (n3);
	\draw[-latex] (n5) edge[bend left=15]  node[midway, sloped, fill=white] {\tiny \texttt{LIKES}} (n8);
	\draw[-latex] (n6) edge[bend right=30] node[midway, sloped, fill=white] {\tiny \texttt{REPLY\_OF}}  (n8);
	\draw[-latex] (n8) edge[bend left=15]  node[inner sep=0.5pt, midway, sloped, fill=white] {\tiny \texttt{HAS\_CREATOR}}  (n5);
	\draw[-latex, draw=Orange, line width=1pt] (n8) edge[out=355, in=315, looseness=7, Orange] node[midway, fill=white, xshift=14pt, inner sep=0pt] {\tiny \texttt{REPLY\_OF}}  (n8);
	
	\draw[>->, Orange, line width=0.5pt, shorten <= 0.05cm, shorten >= 0.05cm] (p) edge node[right] {$m^+$} (phantom);
\end{tikzpicture}
	\caption{A matching of $R$ into the rewritten instance}
	\label{F:rewrite4}
\end{figure}
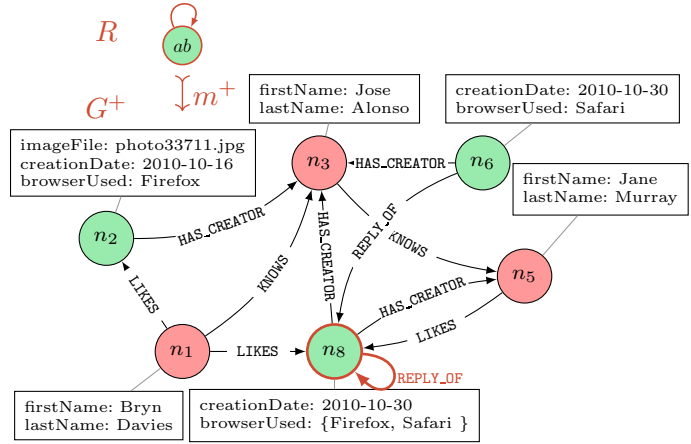

In some use cases, canonical propagation of rewriting to or from the schema does not produce the results we would like. For example, if we clone a node of $S$, we may not wish to clone \emph{every} instance of this node in $G$; instead, we may wish to partition the existing instances in $G$ into those that are now an instance of one clone versus the others that are now an instance of the other clone.

In effect, such a \emph{controlled} propagation from the schema amounts to performing the canonical propagation followed by a `garbage collection' phase where all undesired clones of $G$ are deleted. This requires us to specify, in addition to the rule itself, the instances affected by the garbage collection phase. The case of canonical propagation occurs if we specify no garbage collection. Propagation to the schema can also be controlled; this amounts to specifying which newly-added nodes of $S$ should in fact be merged with pre-existing nodes. A typical use case of controlled propagation to the schema occurs if we wish to propagate a rewrite that adds two node to $G$, only one of which is an instance of an existing node of $S$.

\noindent {\textbf{\texttt{ReGraph} revisited.}}
The \texttt{ReGraph} library enables us to express and apply rewriting rules as above. It also computes all necessary propagation of rewrites automatically so that, given a hierarchy and a rewrite of one of its graphs, it performs all necessary rewrites and reconstructs the updated hierarchy. In the case of an instance-schema hierarchy $h : G \rightarrow S$, this guarantees that any rewrite (of $G$ or $S$) results in a valid instance of the (updated) schema.

Controlled propagation is expressed in \texttt{ReGraph} by specifying a \emph{relation} between the rule and the graph to which we are propagating. In the case above of a partition of the nodes of $G$, this relation would state explicitly which nodes of $G$ correspond to which nodes of the rewritten schema; we will see an example of this in the next section. An analogous relation can be used to specify controlled propagation to the schema.

\section{Schema evolution}\label{S:evol}


In this section, we investigate the use of \texttt{ReGraph}-style rewriting with propagation in our setting where the hierarchy is of the form $h : G \rightarrow S$. We discuss the use of propagation from and to the schema to capture rigorously the distinction between prescriptive PG schema design (and enforcement) and descriptive PG schema evolution and show how this two-way `to-and-from' dialectic can be used to formalize/make explicit the \emph{process} of PG schema development.

\subsection{SMOs, mathematically}

Let us consider the question of providing a language of SMOs from a different perspective. We have defined, in Section \ref{S:DDL}, a DDL in which we can specify schemas and, in Section \ref{S:pp}, we explained how to interpret such a specification as a property graph. In Section \ref{S:PG-rewriting}, we defined graph rewriting rules that we can use to modify a schema. These rewrites can be viewed as a formalization of a class of the SMOs of general interest, that correspond to 
the schema modifications described in Section \ref{S:SML}.

However, in order to exploit this, we need to be able to `read back' the modified schema graph into DDL syntax, \ie we must limit the kinds of mathematical rewrites we perform so that the modified PG schema is still itself the interpretation of some DDL specification. This may not always be possible because, as discussed above, the property graph interpretation of a DDL schema does not represent inheritance explicitly. As such, we could delete a property of the PG schema which, in the DDL schema, came from inheriting some element type. Assuming that other node types of the DDL schema also inherited the same element type, they should all also lose that property---but we have no way to enforce this unless we represent inheritance in our formalism.

In the next section, we explain---in the context of the use case of \emph{concept fine-graining}---how to capture inheritance formally. We then show---for a restricted class of rewrites: restrictive rewrites that \emph{only} clone (not delete) and expansive rewrites that only add (not merge)---that we can rewrite our PG schema and then recover the DDL schema to which it corresponds. This means that, starting from a `before' DDL schema, we obtain an `after' DDL schema. In itself, this does not define a concrete syntax of SMOs---whose job would be precisely to \emph{transform} the `before' into the `after'. However, it does sharply focus our attention on the requirements that the SMO syntax must fulfil.

\subsection{Conceptual fine-graining}

The full SNB schema \cite{ErlingALCGPPB15} contains an abstract class of \cypherinline{Message}s that is inherited by the concrete classes of \cypherinline{Post}s and \cypherinline{Comment}s. We might imagine that, at an earlier stage of its development, the schema contained only the \cypherinline{Message} class but that its users began to evolve an \emph{ad hoc} refinement of this class by adding a new \emph{descriptive} property to instances of \cypherinline{Message}s to specify whether they are intended as a `post' or as a `comment'.
In order to maintain validity of the instance, this property would have had to be explicitly added to the \cypherinline{Message} node of the schema. Eventually, this \emph{ad hoc} evolution of the schema could have been codified \emph{prescriptively} by cloning the \cypherinline{Message} node into \cypherinline{Post} and \cypherinline{Comment} nodes. A major update of the instance would then have been necessary in order to recover validity with respect to this finer-grained schema.


Let us replay this hypothetical scenario in our \texttt{ReGraph}-based framework.
Our starting point $S$ is the schema of Figure~\ref{F:new-schema} with the instance $G$ of Figure~\ref{F:SNB-instance} where all blue and yellow nodes are therefore instances of \cypherinline{Message}.

\vspace{3pt}

\noindent {\textbf{Descriptive updates.}} We begin by defining a rewriting rule that adds a property \cypherinline{type:post} to a node and applying this rule to some message node of the instance. According to canonical propagation of rewriting, this would add the property \cypherinline{type:post} to the \cypherinline{Message} node of the schema. Subsequent applications of this rule to other message nodes would update \emph{only} the instance; \texttt{ReGraph} would not propagate to the schema as the rule respects the updated schema. However, if we create a second rule that adds the property \cypherinline{type:comment} to a node, applying this rule to some message node of the instance \emph{would} induce a second propagation to the schema because a \emph{novel} value is being associated with the property: the overall effect would be to update the property in the schema to \cypherinline{type:\{post,comment\}}. In other words, \texttt{ReGraph}-style propagation \emph{automatically} updates the schema as and when users add such \emph{descriptive} \cypherinline{type} properties to the instance.

\vspace{3pt}

\begin{figure}[h]
	\centering
		\begin{tikzpicture}[person/.style={draw, circle, fill=red!40!white}, message/.style={draw, circle, fill=green!80!blue!40!white}, post/.style={draw, circle, fill=blue!40!white}, comment/.style={draw, circle, fill=yellow!40!white}, attributes/.style={draw},attribute edge/.style={-, black!40!white}, graph edge/.style={-latex}, neutral/.style={draw, circle, fill=black!20!white}]
		\linespread{0.5}
	
	\node[person] (n1) at (0, 0) {\scriptsize $n_1$};
	\node[draw, minimum width=1.5cm, align=left] (n1_attrs) at (-1.1, -0.85) {\tiny firstName:\ Bryn\\ \tiny lastName:\ Davies};	
	
	\node[post] (n2) at (-1, 1.5) {\scriptsize $n_2$};
	\node[draw, minimum width=1.5cm, align=left] (n2_attrs) at (-0.65, 2.5) { \tiny imageFile:\ photo33711.jpg \\ \tiny creationDate:\ 2010-10-16\\  \tiny browserUsed:\ Firefox \\ {\tiny type:\ post}};	
	
	\node[person] (n3) at (1.8, 2.5) {\scriptsize $n_3$};
	\node[draw, minimum width=1.5cm, align=left] (n3_attrs) at (1.5, 3.45) { \tiny firstName:\ Jose\\ \tiny lastName:\ Alonso};	
	
	\node[post, draw=Orange, line width=1pt] (n8) at (1.75, 0) {\scriptsize $n_8$};
	\node[draw, minimum width=1.5cm, align=left] (n8_attrs) at (1.75, -0.85) {
		\tiny creationDate: 2010-10-30\\
		\tiny browserUsed:\\
		\tiny\ \{Firefox, Safari \}};	
	
	\node[comment, draw=Orange, line width=1pt] (n9) at (3, 0.5) {\scriptsize $n_9$};
	\node[draw, minimum width=1.5cm, align=left] (n9_attrs) at (4.5, -0.5) {\tiny creationDate:\\
	\tiny\ \ \ \ 2010-10-30\\ \tiny browserUsed:\\
	\tiny\ \{Firefox, Safari \}};		
	
	\node[person] (n5) at (4.5, 1) {\scriptsize $n_5$};
	\node[draw, minimum width=1.5cm, align=left] (n5_attrs) at (4.5, 1.75) {\tiny firstName:\ Jane\\ \tiny lastName:\ Murray};		
	
	\node[comment] (n6) at (3.95, 2.5) {\scriptsize $n_6$};
	\node[draw, minimum width=1.5cm, align=left] (n6_attrs) at (4.25, 3.35) {\tiny creationDate:\ 2010-10-30\\ \tiny browserUsed:\ Safari \\ {\tiny type:\ comment}};	
	
	
	\draw[-, black!40!white] (n1_attrs) edge (n1);
	\draw[-, black!40!white] (n2_attrs) edge (n2);
	\draw[-, black!40!white] (n3_attrs) edge (n3);
	\draw[-, black!40!white] (n8_attrs) edge (n8);
	\draw[-, black!40!white] (n9_attrs) edge (n9);
	\draw[-, black!40!white] (n6_attrs) edge (n6);	
	\draw[-, black!40!white] (n5_attrs) edge (n5);	
	
	\draw[-latex] (n1) edge node[midway, sloped, fill=white, inner sep=1pt] {\tiny \texttt{LIKES}}  (n2);
	\draw[-latex] (n1) edge[bend right=20] node[midway, sloped, fill=white, inner sep=1pt] {\tiny \texttt{KNOWS}}  (n3);
	\draw[-latex] (n3) edge[bend right=20] node[midway, sloped, fill=white, inner sep=1pt] {\tiny \texttt{KNOWS}}  (n5);
	\draw[-latex] (n1) edge  node[midway, sloped, fill=white, inner sep=1pt] {\tiny \texttt{LIKES}} (n8);
	\draw[-latex] (n1) edge[bend left=10]  node[midway, sloped, fill=white, inner sep=0.8pt, pos=0.35] {\tiny \texttt{LIKES}} (n9);
	\draw[-latex] (n2) edge[bend right=20]  node[midway, sloped, fill=white, inner sep=0.5pt] {\tiny \texttt{HAS\_CREATOR}}  (n3);
	\draw[-latex] (n8) edge node[midway, sloped, fill=white, inner sep=0.5pt] {\tiny \texttt{HAS\_CREATOR}}   (n3);
	\draw[-latex] (n9) edge node[pos=0.4, midway, sloped, fill=white, inner sep=0.5pt] {\tiny \texttt{HAS\_CREATOR}}   (n3);
	\draw[-latex] (n6) edge  node[midway, sloped, fill=white, inner sep=1pt] {\tiny \texttt{HAS\_CREATOR}} (n3);
	\draw[-latex, draw=Orange, line width=1pt] (n9) edge[out=65, in=15, looseness=5, Orange] node[below, fill=white, xshift=13pt, yshift=-10pt, inner sep=0pt] {\tiny \texttt{REPLY\_OF}}  (n9);
	\draw[-latex, draw=Orange, line width=1pt] (n9) edge[Orange] node[below, fill=white, xshift=10pt, yshift=-5pt, inner sep=0pt] {\tiny \texttt{REPLY\_OF}}  (n8);

\end{tikzpicture}
	\caption{The updated instance after controlled propagation}
	\label{F:control2}
\end{figure}
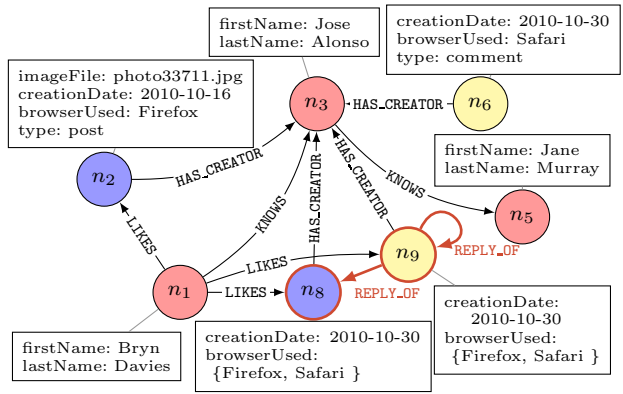

\noindent {\textbf{Prescriptive updates.}} We continue by defining a third rule that clones a node with property \cypherinline{type:\{post,comment\}} and applying it to the \cypherinline{Message} node of the schema; see Figure \ref{F:control1}.
The effect of this is precisely to split the \cypherinline{Message} node into \cypherinline{Post} and \cypherinline{Comment} nodes as in Figure \ref{F:SNB}. However, if we use canonical propagation to the instance graph, to recover schema validation, this would have the (unintended) effect of duplicating \emph{every} \cypherinline{Message} as both a \cypherinline{Post} \emph{and} a \cypherinline{Comment}. Instead, we perform a \emph{controlled} propagation which would map nodes of $G$ with \cypherinline{type:post} to the node \cypherinline{Post} of the updated schema (and similarly for \cypherinline{Comment}s); see Figure \ref{F:control2}.

In other words, \texttt{ReGraph}-style controlled propagation updates the instance after a \emph{prescriptive} update of the schema. Let us note, however, that in order to \emph{specify} the controlled propagation, we first need to perform a \emph{query} on the instance: in this example, we need to match all instances of \cypherinline{Message} with property \cypherinline{type:post} (and similarly for \cypherinline{type:comment})) in order to partition the instance appropriately. In particular, an instance of \cypherinline{Message} that happens not to have the \cypherinline{type} property \emph{will} be cloned as both a \cypherinline{Post} \emph{and} a \cypherinline{Comment}.\\

\begin{figure}[h]
	\centering
		\begin{tikzpicture}[person/.style={draw, circle, fill=red!40!white}, message/.style={draw, circle, fill=green!80!blue!40!white}, post/.style={draw, circle, fill=blue!40!white}, comment/.style={draw, circle, fill=yellow!40!white}, attributes/.style={draw},attribute edge/.style={-, black!40!white}, graph edge/.style={-latex}, neutral/.style={draw, circle, fill=black!20!white}]
		\linespread{0.5}

		\node (l_label) at (-1, 6.25) {\color{Orange} $L$};
  
		\node[scale=0.5, message, draw=Orange, minimum width=1.8cm] (message) at (0, 6) {Message};
		\node[draw, minimum width=1.5cm, align=left] (message_attrs) at (0, 6.85) {\tiny imageFile:\ \texttt{STRING}\\ \tiny \color{darkgreen} \textbf{type:\ \{post,\ comment\}}};
	
		\draw[-, black!40!white] (message_attrs) edge (message);
	
		\draw[-latex, Orange] (message) edge[out=25,in=335, looseness=5] (message);
	
		\node (l_phantom_east) at (1, 5.75) {};
		\node (l_phantom_south) at (0, 5.25) {};
		
		\node (p_label) at (2.5, 6.25) {\color{Orange} $P$};
		
		\node[scale=0.5, post, draw=Orange, minimum width=1.8cm] (post) at (3.25, 6) {Post};
		\node[draw, minimum width=1.5cm, align=left] (post_attrs) at (3.25, 6.85) {\tiny imageFile:\ \texttt{STRING} \\ \tiny \color{darkgreen} \textbf{type:\ post}};
	
		\node[scale=0.5,  comment, draw=Orange, minimum width=1.8cm] (comment) at (5.35, 6) {Comment};
		\node[draw, minimum width=1.5cm, align=left] (comment_attrs) at (5.55, 6.85) {\tiny \color{darkgreen} \textbf{type:\ comment}};

		\draw[-, black!40!white] (post_attrs) edge (post);
		\draw[-, black!40!white] (comment_attrs) edge (comment);

		\draw[-latex, Orange] (comment) edge[bend right=15] (post);
		\draw[-latex, Orange] (comment) edge[out=30,in=-15, looseness=5] (comment);
		
		\node (p_phantom) at (2.75, 5.75) {};

		\node (s_label) at (-1.5, 4.65) {\color{Orange} $S$};
		\node[scale=0.5] (schema) at (0, 4.65) {
			\begin{tikzpicture}
			\node[person, minimum width=1.5cm] (person) at (0, 0) {\scriptsize Person};	
			\node[message, minimum width=1.5cm, draw=Orange, line width=1pt] (message) at (3, 0) {\scriptsize Message};
	
			\draw[-latex] (person) edge[out=115,in=65,looseness=5] (person);
			\draw[-latex] (message) edge[bend right=15] (person);
			\draw[-latex] (person) edge[bend right=15] (message);

			\draw[-latex, Orange, line width=1pt] (message) edge[out=115,in=65, looseness=5] (message);

		\end{tikzpicture}
	};

	\node (g_label) at (-1.5, 3.35) {\color{Orange} $G$};
	\node (phantom) at (0, 3) {};
	
	\node[person] (n1) at (0, 0) {\scriptsize $n_1$};
	\node[draw=black!40!white, minimum width=1.5cm, align=left] (n1_attrs) at (-1.1, -0.85) {\color{black!40!white} \tiny firstName:\ Bryn\\ \color{black!40!white} \tiny lastName:\ Davies};	
	
	\node[message] (n2) at (-1, 1.5) {\scriptsize $n_2$};
	\node[draw=black!40!white, minimum width=1.5cm, align=left] (n2_attrs) at (-0.85, 2.5) {\color{black!40!white} \tiny imageFile:\ photo33711.jpg \\ \color{black!40!white} \tiny creationDate:\ 2010-10-16\\ \color{black!40!white}  \tiny browserUsed:\ Firefox \\ {\tiny \color{darkgreen} \textbf{type:\ post}}};	
	
	\node[person] (n3) at (1.8, 2.5) {\scriptsize $n_3$};
	\node[draw=black!40!white, minimum width=1.5cm, align=left] (n3_attrs) at (1.5, 3.5) {\color{black!40!white} \tiny firstName:\ Jose\\ \color{black!40!white} \tiny lastName:\ Alonso};	
	
	\node[message, draw=Orange, line width=1pt] (n8) at (2, 0) {\scriptsize $n_8$};
	\node[draw=black!40!white, minimum width=1.5cm, align=left] (n8_attrs) at (2, -0.85) {\color{black!40!white} \tiny creationDate:\ 2010-10-30\\ \color{black!40!white} \tiny browserUsed:\ \{Firefox, Safari \}};		
	
	\node[person] (n5) at (4.5, 1) {\scriptsize $n_5$};
	\node[draw=black!40!white, minimum width=1.5cm, align=left] (n5_attrs) at (4.5, 1.75) {\color{black!40!white} \tiny firstName:\ Jane\\ \color{black!40!white} \tiny lastName:\ Murray};		
	
	\node[message] (n6) at (4, 2.5) {\scriptsize $n_6$};
	\node[draw=black!40!white, minimum width=1.5cm, align=left] (n6_attrs) at (4.25, 3.5) {\color{black!40!white} \tiny creationDate:\ 2010-10-30\\ \color{black!40!white} \tiny browserUsed:\ Safari \\ \color{black!40!white} {\tiny \color{darkgreen} \textbf{type:\ comment}}};	
	
	
	\draw[-, black!40!white] (n1_attrs) edge (n1);
	\draw[-, black!40!white] (n2_attrs) edge (n2);
	\draw[-, black!40!white] (n3_attrs) edge (n3);
	\draw[-, black!40!white] (n8_attrs) edge (n8);
	\draw[-, black!40!white] (n6_attrs) edge (n6);	
	\draw[-, black!40!white] (n5_attrs) edge (n5);	
	
	\draw[-latex] (n1) edge node[midway, sloped, fill=white, inner sep=1pt] {\tiny \texttt{LIKES}}  (n2);
	\draw[-latex] (n1) edge[bend right=20] node[midway, sloped, fill=white, inner sep=1pt] {\tiny \texttt{KNOWS}}  (n3);
	\draw[-latex] (n3) edge[bend right=20] node[midway, sloped, fill=white, inner sep=1pt] {\tiny \texttt{KNOWS}}  (n5);
	\draw[-latex] (n1) edge  node[midway, sloped, fill=white, inner sep=1pt] {\tiny \texttt{LIKES}} (n8);
	\draw[-latex] (n2) edge[bend right=20]  node[midway, sloped, fill=white, inner sep=1pt] {\tiny \texttt{HAS\_CREATOR}}  (n3);
	\draw[-latex] (n8) edge node[midway, sloped, fill=white, inner sep=1pt] {\tiny \texttt{HAS\_CREATOR}}   (n3);
	\draw[-latex] (n6) edge  node[midway, sloped, fill=white, inner sep=1pt] {\tiny \texttt{HAS\_CREATOR}} (n3);
	\draw[-latex, draw=Orange, line width=1pt] (n8) edge[out=355, in=315, looseness=7, Orange] node[midway, fill=white, xshift=13pt, inner sep=0pt] {\tiny \texttt{REPLY\_OF}}  (n8);
	
	\draw[>->, Orange, line width=0.5pt, shorten <= 0.05cm, shorten >= -0.5cm] (l_phantom_south) edge node[pos=1, right, yshift=-7pt] {\scriptsize $m$} (schema);
	\draw[->, Orange, line width=0.5pt, shorten <= 0.05cm, shorten >= 0.05cm] (phantom) edge node[right] {\scriptsize $h$} (schema);
	\draw[->, Orange, line width=0.5pt, shorten <= 0.08cm, shorten >= 0.09cm] (p_phantom) edge node[above] {\scriptsize $l^-$} (l_phantom_east);
	
	\draw[darkgreen, line width=1pt, dashed] (post) edge[bend left=25] (n2);
	\draw[darkgreen, line width=1pt, dashed] (comment) edge[bend left] (n6);
	
\end{tikzpicture}
	\caption{Prescriptive update of the schema with controlled propagation expressed by the dashed lines}
	\label{F:control1}
\end{figure}

Overall, this example shows how the functionality provided by \texttt{ReGraph} to propagate changes automatically to \emph{and} from the schema provides strong support for this use case---that combines prescriptive and descriptive aspects---which is a typical example of what can happen during the development of a system from an early agile state to a late mature state.

\subsection{Towards SMOs}

Note that the prescriptive update that clones the \cypherinline{Message} node corresponds precisely to (i) defining two new node types that, in the DDL syntax, inherit \cypherinline{Message}; and (ii) deleting the \cypherinline{Message} node type. This suggests that we can keep track of inheritance in our formalism by maintaining not only the \emph{current} schema but also the rule applications that we used to construct it, \ie an \emph{audit trail} that recapitulates the schema development process. This implies that all properties added to a node must be \emph{pushed through} to the end of the trail so all inheriting nodes also get the new property.

\setlength{\columnsep}{10pt}
\begin{wrapfigure}[5]{r}{2cm}
	\vspace{-10pt}
	{\small
	\begin{tikzcd}[row sep = tiny]
	& S^- \arrow[dl] \\
	S \arrow[dr] \\
	& S^+
	\end{tikzcd}}
\end{wrapfigure}
More precisely, suppose we start with a schema $S$ and (i) specify some inheritance/cloning, through a restrictive rewrite, leading to a refined schema $S^-$; and also (ii) specify some extension/addition of properties, through a (direct or propagated) expansive rewrite, leading to a richer schema $S^+$.

\setlength{\columnsep}{10pt}
\begin{wrapfigure}[5]{r}{2cm}
	\vspace{-5pt}
	\hspace{-35pt}
	{\small
	\begin{tikzcd}[row sep = tiny]
		& S^- \arrow[dr] \\
		& & S' \arrow[dl] \\
		& S^+
	\end{tikzcd}}
\end{wrapfigure}
We can \emph{push} the expansive rewrite through to $S^-$ to obtain a new schema $S'$ that combines both updates, \ie if rewrite (ii) added a property to a node that was cloned by rewrite (i), all the clones would have the new property in $S'$.

As such, $S'$---but neither $S^-$ nor $S^+$---can be read back to a DDL schema which is a natural modification of the original DDL schema (for $S$): it adds and removes the necessary node types and adds the necessary properties to the appropriate element types. In a sense, this means that we \emph{could} define this class of schema modifications by literally importing the initial DDL schema into \texttt{ReGraph}, rewriting it \emph{in situ} then reading back the updated DDL schema---but, of course, this is not a practical proposal. Nonetheless, our analysis tells us that the only SMOs needed to express the inherit and extension operations performed by this restricted class of rewriting rules are the \textbf{create}, \textbf{change} and \textbf{split} operations from Section \ref{S:SML}.

We must leave to future work the extension of our analysis to general rewriting rules and the definition of a practical concrete syntax for SMOs.

Let us conclude this Section by noting that, somehow dual to additions, deletions in an audit trail must be pushed back to the beginning as nodes cannot pick and choose which properties they inherit from an element type. On the other hand, deleting a property in the instance graph would not propagate to the schema and would require no special treatment. 

\section{Implementation}\label{S:implem}

In this section we would like to address the implementation issues  that one must tackle in order to build the system presented above using a generic PG query engine. Although our current implementation (a prototype system proposed by \texttt{ReGraph}) is built on the Cypher language, we have opted to present the graph pattern matching and update operations in a generic fashion and to have operations that are agnostic to the particular choice of query language. In fact, any query language can be enhanced with the additional data manipulation capabilities that are needed in order to implement graph rewriting and propagation.

It should be noted that in this paper we do not address the problem of performance and scalability of the proposed system, which would heavily rely on such of the underlying Cypher query engine. Our main goal is to stress the conciseness and comprehensibility of our two-way schema/instance updates, as well as the possibility to automate these updates saving the user from thousands of lines of routine queries necessary to perform schema/instance rewriting and propagation.

\subsection{Property graph rewriting}

Given a property graph $G$, a rewriting rule $L \leftarrow P \rightarrow R$ and an instance $m: L \rightarrowtail G$ we would like to obtain a graph query that performs in-place rewriting of $G$. Such a query would consist of a \emph{graph pattern matching} subquery along with an \emph{update} subquery which, when executed, would transform $G$ into $G^+$. 

\vspace{3pt}

\noindent {\textbf{Graph pattern matching.}} Graph pattern matching can be performed using a typical match clause in a given graph query language. It should allow us to obtain a set of bindings from the nodes and edges of $L$ to the actual nodes and edges of $G$. Note that, in our setting, the instance of a rule is always given by an \emph{injective} homomorphism $m$ which means that the nodes and edges of our patterns are always distinct graph objects. Such semantics is also known as isomorphism-based semantics in the literature of graph query languages \cite{AnglesABHRV17} where both node and edge variables must be mapped one-to-one. 

\vspace{3pt}

\noindent {\textbf{Update subqueries.}} Let $G$ be a property graph as defined in Section \ref{S:pp}. We now show how to formulate the elementary graph transformations needed for graph rewriting and we do so by considering the aforementioned ``agnostic'' operations such as addition/removal of graph elements and their properties. 

\vspace{3pt}

Some primitive graph transformations have immediate counterparts in most modern graph query languages (such as addition/removal of nodes and edges and set-up of properties). However, it is usually necessary to \emph{program} clone and merge operations as discussed in the remainder of this section. 

To ease our presentation we first define the notion of a \emph{property dictionary} associated with a node or edge as well as the operation of \emph{dictionary union}. We define the set of property keys associated to a graph element $x \in N \cup E$ as $\op{keys}(x) = \{ k \in \Keys | (x, k) \in P\}$ and, for every such key $k \in \op{keys}(x)$, the set of its property values as $\op{vals}(x, k) = \{ v \in \Nval | ((x, k), v) \in \nu \}$. We can now define a property dictionary of an element $x$ as $\op{dict}(x) = \{ (k, \op{vals}(x, k)) | k \in \op{keys}(x) \}$. As such, any dictionary is an element of $\Keys \times 2^{\Nval}$. 

The operation of \emph{dictionary union} $d_1 \cup d_2$ for any two dictionaries $d_1, d_2 \in \Keys \times 2^{\Nval}$ can be now defined as the union of their keys and the value sets corresponding to these keys, where if a particular key $k$ is present in, for example, $d_1$ but not in $d_2$, the set of values corresponding to $k$ in the union $d_1 \cup d_2$ is simply the value set of $k$ from $d_1$. 

\vspace{3pt}
\noindent {\textbf{Clone.}} The cloning operation for a particular node of interest creates a new node and copies to it all the properties of the original one. All the incident edges, together with their properties, are also copied to the newly created node. Query \ref{algo:clone_pseudo_query} illustrates such a cloning operation expressed as a language-agnostic pseudo-query; while Appendix \ref{App:clone} contains a real Cypher 9 query generated by the prototype system implemented in \texttt{ReGraph}.

\vspace{3pt}
\noindent {\textbf{Merge.}} The merging operation for a given set of nodes can be implemented in two different ways. First, we can create a new node (that we will use as the result of merge), set its property dictionary to be the union of the dictionaries of all the nodes we merge, reconnect all the neighbours and finally remove all the nodes we are merging. The second version of merge can be implemented by picking an arbitrary node from the set of nodes that need to be merged and using it, instead of a new node, to perform the above operations. In our implementation, we chose the second version as it lets us save some extra operations of property and edge addition. 

There is, however, a subtle issue to be discussed concerning the simplicity of graphs in which we want to perform a merging operation. Its behaviour, when reconnecting edges, differs slightly for simple and non-simple graphs. This is due to the fact that if two nodes to be merged have edges from/to the same neighbour in a simple graph, these edges should be merged into a single edge
whose properties are given by the union of the original property dictionaries. Therefore, a merge operation in simple graphs has an overhead compared with non-simple graphs.

The pseudo-query for non-simple graphs is illustrated in Query \ref{algo:merge_pseudo_query_non_simple}, while the query for simple ones can be found in Appendix \ref{App:merge_pseudo_query_simple}. Although we focus on simple graphs in this paper, we report both versions in order to show the extra operations required when switching from non-simple graphs to simple graphs. Appendix \ref{App:merge} contains a real Cypher 9 query for merging in simple graphs generated by our prototype.

\begin{pseudoquery}
  \KwData{$G=(N,E,\eta,P,\nu,M)$, $n \in N$ (node to clone)}
  
  create node $n'$
 
  set properties of $n'$ to $\op{dict}(n)$
  
  \For{$s \in successors(n)$}{
  		create edge $e' = (n', s)$ \\
  		set properties of $e' = \op{dict}(e)$, where $e = (n, s)$
  }
  
  \For{$p \in predecessors(n)$}{
  		create edge $e' = (p, n')$ \\
  		set properties of $e' = \op{dict}(e)$, where $e = (p, n)$
  }
 
  \caption{Clone of a node}\label{algo:clone_pseudo_query}
\end{pseudoquery}

\begin{pseudoquery}

  \KwData{$G=(N,E,\eta,P,\nu,M)$, $n, m \in N$ (nodes to merge)}
  
  set properties of $n$ to $\op{dict}(n) \cup \op{dict}(m)$
  
  \For{$s \in successors(m)$}{
  		create edge $e' = (n, s)$ \\
  		set properties of $e' = \op{dict}(e)$, where $e = (m, s)$
  }

  \For{$p \in (predecessors(m)$}{
  		create edge $e' = (p, n)$ \\
  		set properties of $e' = \op{dict}(e)$, where $e = (p, m)$
  }
  
  detach and delete $m$ \\
  
  \vspace{10pt}
  \caption{Merge of two nodes in non-simple graphs}\label{algo:merge_pseudo_query_non_simple}
\end{pseudoquery}

In what follows, for the sake of conciseness, we refer to clone and merge operations as ``clone $n$ into $\{n_1, n_2, \ldots, n_k\}$'' and ``merge $\{n_1, n_2, \ldots, n_k \}$ into $n$'', \ie generalizing clone and merge to work on $k \geq 1$ nodes. They also represent the primitive operations whose incorporation into a graph query language would significantly facilitate the support for graph rewriting and propagation. 


\subsection{Propagation of rewriting}

Let $G$ be an instance graph typed by a schema graph $S$ through an homomorphism $h: G \to S$. In this section, we focus on the case where rewriting of $G$ or $S$ requires propagation of changes (as described in Section \ref{S:PG-rewriting}) and, in particular, how to express this propagation with elementary graph operations that can be translated into typical graph query language clauses. Here we formulate updates of both schema and instance graphs using generic graph update operations as in the previous subsection. Note that, for simplicity of exposition, we assume that both $G$ and $S$ are simple graphs. 

According to our approach for both rewriting cases (rewriting of $S$ to $S^-$ with a restrictive rule and $G$ to $G^+$ with an expansive rule) we first perform rewriting that invalidates $h$, then we compute a set of transformations that ``repair'' the typing by producing a new graph homomorphism. We discuss both cases in the following. 

\vspace{3pt}
\noindent {\textbf{Propagation to instance.}} Recall that a restrictive rewriting of $S$ to $S^-$ gives us the homomorphism $l^-: S^- \to S$ (as in Section \ref{S:PG-rewriting}). Given $h$ and $l^-$ we can construct the following set of pairs 
$h^r = \{ (n, t) \in N_G \times N_{S^-}\ |\ h_{\mathcal{N}}(n) = l^{-}(t)\}$.
This set of pairs can be used to infer the ``repair'' transformations necessary for rewriting of $G$ to $G^-$ and restoring $h^-: G^- \to S^-$. For every node $n \in N_G$ of the instance graph $G$ we define the typing relation with nodes from $S^-$ as $\op{type}(n) = \{ t \in N_{S^-}\ |\ (n, t) \in h^r \}$. Now we can use the following pseudo-query to perform the necessary clones and deletes in $G$ to produce the homomorphism $h_{N}^- : N_{G^-} \to N_{S^-} $ which, due to the fact that our graphs are simple, uniquely defines a homomorphism $h^-: G^- \to S^-$.

\begin{pseudoquery} 
  \For{$n \in N_G$ such that $\op{type}(n) = \emptyset$}{
  	detach delete $n$ \\
  }

  \For{$n \in N_G$ such that $|\op{type}(n)| \geq 1$}{ 	
  	pick an element $t_0 \in \op{type}(n)$ \\	
  	set $h^-(n) = t_0$ \\
  	\For{$t \in \op{type}(n) \setminus \{t_0\}$}{	
  		clone $n$ as $\{ n' \}$ \\
  		set $h^-(n') = t$ \\
  	}
  }
  
  \For{$(s, t) \in E_G$ such that $(h^-(s), h^-(t)) \notin E_{S^-}$}{
  	 delete edge $(s, t)$ \\
  }
  \caption{Propagation to the instance graph}\label{query:propagate_remove_node_merge}
\end{pseudoquery}

\vspace{3pt}
\noindent {\textbf{Propagation to schema.}}We now consider the case where an expansive rewrite of the instance graph $G$ to $G^+$ induces some changes to the schema. Such a rewriting gives us the homomorphism $r^+: G \to G^+$ (as in Section \ref{S:PG-rewriting}). Now given $h$ and $r^+$ we can construct the following set of pairs $h^e = \{ (r^+(n), h(n))\ |\ n \in N_G ) \}$.
In this case $h^e$ can be used to infer the necessary rewriting of the schema graph $S$ to $S^+$ and restoring $h^+: G^+ \to S^+$. For every node $n \in N_{G^+}$ of the rewritten instance graph $G^+$, we define the typing relation with nodes from $S$ as $\op{type}(n) = \{ t \in N_{S}\ |\ (n, t) \in h^e \}$. Query \ref{query:propagate_add_node_merge} performs the node additions and merges followed by necessary edge additions that are required to construct $S^+$ and $h^+: G^+ \to S^+$: 

\begin{pseudoquery} 

  \For{$n \in N_{G^+}$ such that $\op{type}(n) = \emptyset$}{
  	create node $t$ in $S$ \\
  	set $h^+(n) = t$ \\
  }

  \For{$n \in N_{G^+}$ such that $|\op{type}(n)| \geq 1$}{
  	merge $\op{type}(n)$ as $t$ \\
  	set $h^+(n) = t$ \\  	
  }
  \For{$(s, t) \in E_{G^+}$ such that $(h^+(s), h^+(t)) \notin E_{S^+}$}{
  	 create edge $(h^+(s), h^+(t))$ in $S$ \\
  }
  \caption{Propagation to the schema graph}\label{query:propagate_add_node_merge}
\end{pseudoquery}

\section{Discussion}\label{S:discussion}

In this section, we discuss additional details of the implementation of our mathematical framework and its suitability for understanding the requirements of PG schema validation and evolution. 

The Python library \texttt{ReGraph} on which we build was originally based on in-memory \texttt{NetworkX}\footnote{\url{https://networkx.github.io}} and later extended to work directly with the Neo4j graph database. This necessitates a certain amount of encoding in order to (i) maintain the multiple graphs that constitute a hierarchy within the single graph (and so namespace) currently provided by Neo4j; and (ii) to represent the homomorphisms of the hierarchy as edges with a `reserved' label that encodes the typing. Moreover, because we consider properties that have \emph{sets} of values, we must encode this using Neo4j lists. However, we anticipate that future versions of Neo4j will reduce this overhead of encoding, notably by providing native support for multiple graphs.

The encoding into Neo4j allows us to represent rewriting rules as Cypher queries that are computed automatically. Although the current version of Cypher supports basic update operations\footnote{Precisely, in our development we focus on the following Cypher operations: \cypherinline{create} to create nodes and edges, \cypherinline{merge} to match and create edges, \cypherinline{set} to set the properties of nodes and edges, \cypherinline{delete} to delete nodes and edges, \cypherinline{detach delete} to force removal of nodes with incident edges.}, the lack of native support for clone and merge operations leads to a significant blow-up in the size of the query (as reported in Appendix \ref{App:clone} and Appendix \ref{App:merge}), compared to the rule itself, notably due to the requirement that the homomorphic mapping from the instance to the schema must be maintained at all times. A further minor issue arises, in our encoding of merge operations, where we are obliged to make one limited use of the \texttt{apoc} library because we cannot represent a key as a variable. Presumably, one could envision the addition of clone and merge operations, in the style we showed in this paper, to future versions of graph query languages (including GQL, OpenCypher and G-CORE \cite{FrancisGGLLMPRS18, AnglesABBFGLPPS18}.

Finally, let us note that our current interpretation of the semantics of Cypher 9's update operations used in our implementation is based on their practical usage due to the lack of formal semantics for these operations. The formalization of the update fragment of Cypher is actually ongoing\footnote{Leonid Libkin, private communication.} and will soon lead to a formal interpretation, similar to that realized for its read-only fragment \cite{FrancisGGLLMPRS18}. 

An in-depth study of the computational complexity of the schema propagation operations based on graph rewriting rules as presented here falls beyond the scope of our paper. Deciding well-typedness of a graph pattern with no ISA edges under a schema graph
with no implicit object class nodes in the GDM model (roughly corresponding to our DDL) is in PTIME \cite{AnglesG08}, thus leading us to the conjecture that our schema validation is in PTIME as well. However, the precise complexity of schema evolution under our DDL, thus entailing revalidation of entire graph patterns after schema modifications or after instance modifications is unknown and left as one of the open theoretical questions of our work.


Conceptually, in \texttt{ReGraph} we represent a schema as a property graph that contains no real data but only constraints on the data that is permitted. In our development, we provide two possible ways to build and maintain the schema graph $S$: one uses symbolic types to constrain the values that a property can take; the other accumulates its set of permitted values. The first choice more closely matches the PG DDL specification---and usual intuition---of schemas whereas the second has a non-standard flavour of mixing actual data with constraints. Nonetheless, the second option may be of interest, at least in earlier \emph{descriptive} phases of schema development, as it can exploit propagation from an instance to the schema to accumulate sets of permitted values automatically. At some point, more generic constraints on data should start to become apparent and one can switch to the more traditional mode of \emph{prescriptive} schema development using symbolic types.

Our current implementation is external to Neo4j in the sense that schema validation and evolution only make sense through the lens of \texttt{ReGraph}. Although this implies a certain overhead, as discussed above, it nonetheless provides a very useful test-bed for a thorough conceptual and technical debugging of the requirements on a modern, flexible system for PG schema validation and evolution---prior to the significant technical effort that would be required to internalize this into native support for schemas in Neo4j.

Moreover, although we have focused exclusively on schemas and instances in this paper, the external framework provided by \texttt{ReGraph} enables an entirely different mode of use of Neo4j (or other graph databases) which decouples the user from the concrete data model and allows them to define their own domain-specific knowledge representation system. Apart from that, one can work on intermediate representations in between concrete graph instances and schemas, such as updatable graph views or graph summaries. 


\section{Related Work}

Schema evolution~\cite{Rahm2006} is a well established topic in data management. A set of principles ruling out schema and instance evolution under schema constraints were discussed in Hartung et al.~\cite{HartungTR11}.
There are various approaches to increase comfort and efficiency, \eg defining a schema evolution aware query language~\cite{Roddick1992} or by providing a general framework to describe database evolution in the context of evolving applications~\cite{Dominguez2008}.

The Meta Model Management 2.0~\cite{Bernstein2007} of Bernstein et al.\ introduced a comprehensive tooling to match, merge and diff given relational schema versions.
The resulting mappings couple the evolution of both the schema and the data; however, these mappings are complex relationships between heterogeneous schemas, as in data integration and ETL scenarios, \ie they only deal with schema evolution after the fact.


Currently, PRISM~\cite{Curino2008,Moon2009,Curino2013} and InVerDa~\cite{Herrmann2018a,Herrmann2017b} seem to provide the most advanced database schema evolution tools.
PRISM focuses on plain database evolution but also allows the answering of queries using former schema versions with respect to the current data.
InVerDa provides fully co-existing schema versions via bidirectional transformations~\cite{Terwilliger2012} with symmetric relational lenses~\cite{Hofmann2011}.

Another interesting category of tools targets the co-evolution of different artefacts in an information system, \eg MoDEF~\cite{Terwilliger2010} introduces an IDE extension to automate the co-evolution of the evolving client schemas and the store.
However, none of these approaches steps away from the underlying assumption of a prescriptive schema.

Apart from relational databases, schema evolution is a hot topic for XML databases and ontology management systems, as surveyed in~\cite{HartungTR11}. 
The major data vendors, such as Oracle, Microsoft SQL Server, and IBM DB2, offer support for XML Schema. Other research efforts, such as X-Evolution~\cite{MesitiCSG06} and XEM~\cite{KramerSCCR01}, addressed the problem of incremental XML validation, where incremental means that empty or default XML elements are often inserted to fill the gaps where a XML document no longer validates. Due to the tree-shaped nature of XML data, these approaches are quite different from ours and are still focused on prescriptive schema---with the exception of XML schema  of type \texttt{xsd:anyType} which can encode unconstrained XML content.  

SHACL~\cite{shacl}  is a language for validating RDF graphs against a set of conditions. These conditions are provided as shapes under the form of an RDF graph. Shapes are used to validate RDF instances against a set of conditions and they can also be viewed as descriptors of the data that do satisfy these conditions. SHACL supports RDF terms restrictions (e.g. value restrictions, allowed values, datatypes comparison etc), cardinality constraints, and predicate constraints (e.g. required predicates, class-specific property range etc). 

Ontologies are conceptually more abstract models than database schemas and range from controlled vocabularies and thesauri over is-a hierarchies/taxonomies and directed acyclic graphs~\cite{HartungTR11}.
Instances have different roles in ontology management systems and typically lie in completely separate data sources. Moreover, ontologies are usually representatives of a specific domain and are the final outcome of collaborative editing from one or more domain experts. 

Research on ontologies also considered the problem of update propagation to instances using Description Logic mappings~\cite{KharlamovZC13,WuL14,WangWZQ15}. However, the ontology formalisms are confined to interpretation from a restricted set of experts as opposed to DDL-like languages in RDBMS and in particular to the DDL proposed in this paper for graph databases. Description Logic mappings are also  quite complex when contrasted with the implicit homomorphisms considered in our work. 

The distinction between descriptive and prescriptive schemas as carried out in our paper is reminiscent of open and close tuple types as 
used for instance in JSON~\cite{OngPV14}. An open type allows a tuple to contain additional attributes beyond those appearing in the schema declaration, whereas a closed type would not allow it. However, the schema flexibility pointed out in our work affects not only types but entire portions of the 
schemas and as such is more general.

Graph rewriting has been used in a variety of areas related to knowledge representation and meta-modelling. For example, triple graph grammars \cite{Schurr1994,Konigs2006,Anjorin2015}---which correspond very closely to our rewriting rules---provide a means to specify bidirectional model transformations and have been used in various applications such as conformance testing and model synchronization.
Another example is the \texttt{KAMI} bio-curation tool \cite{harmer2017bio}, which represents (i) knowledge about protein--protein interactions as graphs; and (ii) updates of knowledge as rewriting rules that propagate to a model-specific schema in descriptive fashion; and moreover (iii) provides a fixed, prescriptive meta-model that constrains the entire system. As such, we see---albeit in a three-level rather than two-level system---an example of the co-existence of more or less mature aspects of schemas within a single application which enables the tool to remain responsive to novel knowledge---provided that that knowledge at least fits within its view of the universe, as defined by its meta-model.

\section{Concluding Remarks}

We have presented a schema DDL for property graphs following the ASCII-art syntax inspired by Cypher. We have shown how schema validation and schema evolution for graphs can be simulated via a mathematical framework that allows to enforce schema and express propagation from schema to instance and vice versa. 
We have discussed how to achieve modern schema requirements for property graph databases by offering support for both prescriptive and descriptive schemas. 
We have discussed an implementation in a pseudo-query language, which is agnostic to concrete graph query language syntax, and provided some details and discussion of our specific encoding in Cypher 9.

We believe that our work can be extended in at least two possible directions. The first direction would add a third layer to the graph hierarchy and study how to apply modifications to a hybrid, or summary, graph $V$ that lies between the instance and the schema, \ie in a hierarchy of the form $G \rightarrow V \rightarrow S$, and which would play the role of an updatable graph view.
Secondly, the preliminary discussion in this paper concerning a schema manipulation language would require the study and definition of concrete syntax proposals for such a language. We hope that this work already provides insights towards that goal and triggers a discussion on these languages---that are much needed for application development, from prototyping to production.


\pagebreak

\section*{Appendix}
\appendix
\section{LDBC SNB schema}

The entire DDL schema encoding of the LDBC SNB benchmark is reported below:\\

\begin{lstlisting}
CREATE GRAPH TYPE snb (
  Person { 
  	creationDate  : TIMESTAMP,
    firstName     : STRING,
    lastName      : STRING,
    gender        : STRING,
    birthday      : DATE,
    email         : STRING,
    speaks        : STRING,
    browserUsed   : STRING,
    locationIP    : STRING
  },
  Organisation {
    name : STRING, url : STRING
  },
  Company <: Organisation {},
  University <: Organisation {},
  Message {
    creationDate  : TIMESTAMP,
    browserUsed   : STRING,
    locationIP    : STRING,
    content       : STRING?,
    length        : INTEGER
  },
  Comment <: Message {},
  Post <: Message {
    language : STRING?, imageFile : STRING?
  },
  Forum {
    title : STRING, creationDate : TIMESTAMP
  },
  TagClass {
    name : STRING, url : STRING
  },
  Place {
    name : STRING, url : STRING
  },
  City <: Place {},
  Continent <: Place {},
  Country <: Place {},
  Tag {
    name : STRING, url : STRING
  },
  HAS_TYPE {},
  HAS_TAG {},
  IS_SUBCLASS_OF {},
  HAS_MODERATOR {},
  HAS_CREATOR {},
  REPLY_OF {},
  HAS_INTEREST {},
  CONTAINER_OF {},
  IS_PART_OF {},
  IS_LOCATED_IN {},
  KNOWS {
    creationDate : TIMESTAMP
  },
  HAS_MEMBER {
    joinDate : TIMESTAMP
  },
  WORK_AT {
    workFrom : INTEGER
  },
  STUDY_AT {
    classYear : INTEGER
  },
  LIKES {
    creationDate : TIMESTAMP
  },
  (Post),
  (Comment),
  (Continent),
  (Country),
  (City),
  (University),
  (Company),
  (Tag),
  (Person),
  (Forum),
  (TagClass),
  (Country)-[IS_PART_OF]->(Continent),
  (Forum)-[HAS_TAG]->(Tag),
  (Person)-[IS_LOCATED_IN]->(City),
  (Comment)-[REPLY_OF]->(Message),
  (University)-[IS_LOCATED_IN]->(City),
  (Person)-[HAS_INTEREST]->(Tag),
  (TagClass)-[IS_SUBCLASS_OF]->(TagClass),
  (City)-[IS_PART_OF]-><1>(Country),
  (Person)-[WORK_AT]->(Company),
  (Forum)-[HAS_MODERATOR]->(Person),
  (Forum)-[HAS_MEMBER]->(Person),
  (Message)-[HAS_CREATOR]->(Person),
  (Tag)-[HAS_TYPE]->(TagClass),
  (Company)-[IS_LOCATED_IN]->(Country),
  (Message)-[HAS_TAG]->(Tag),
  (Message)-[IS_LOCATED_IN]->(Country),
  (Person)-[STUDY_AT]->(University),
  (Person)-[LIKES]->(Message),
  (Forum)-[CONTAINER_OF]->(Post),
  (Person)-[KNOWS]->(Person)
)
\end{lstlisting}


\section{Merging pseudoquery for simple graphs}\label{App:merge_pseudo_query_simple}
\begin{pseudoquery}
  \KwData{$G=(N,E,\eta,P,\nu,M)$, $n, m \in N$ (nodes to merge)}
  
  \emph{// here we pick $n$ to serve as the result of merge} \\
  set properties of $n$ to $\op{dict}(n) \cup \op{dict}(m)$
  
  \For{$s \in successors(m)\setminus\{n, m\}$ such that $s \notin successors(n)$}{
  		create edge $e' = (n, s)$ \\
  		set properties of $e' = \op{dict}(e)$, where $e = (m, s)$
  }
  
  \For{$s \in successors(m)\setminus\{n, m\}$ such that $s \in successors(n)$}{
  		set properties of $e' = \op{dict}(e) \cup \op{dict}(e')$, where $e = (m, s)$ and $e' = (n, s)$
  }
  
  \For{$p \in predecessors(m)\setminus\{n, m\}$ such that $p \notin predecessors(n)$}{
  		create edge $e' = (p, n)$ \\
  		set properties of $e' = \op{dict}(e)$, where $e = (p, m)$
  }
  
  \For{$p \in predecessors(m)\setminus\{n, m\}$ such that $p \in predecessors(n)$}{
  		set properties of $e' = \op{dict}(e) \cup \op{dict}(e')$, where $e = (p, m)$ and $e' = (p, n)$
  }
  
  $l := (n, n)$ \\
  $L := \{e \in inedges(m)\ such\ that\ source(e) \in \{n, m\}\}  \cup \{e \in outedges(m)\ such\ that\ target(e) \in \{n, m\}\}$ \\
 
  \uIf{$l \notin E$ and $L \neq \emptyset$}
  {
    create edge $l$
  }
 
  set properties of $l = \bigcup_{l' \in loops}{\op{dict}(l')}$ \\
  
  detach and delete $m$ \\
  
  \vspace{10pt}
\end{pseudoquery}

\section{Clone Cypher query}\label{App:clone}

\begin{lstlisting}

// Query performing clone of a node
MATCH (a { id : 'a' })
// create a node corresponding to the clone
CREATE (a1) 
WITH a, a1
SET a1 = a
WITH a, a1

// match successors and out-edges
OPTIONAL MATCH (a)-[out_edge:edge]->(suc)
WITH a, a1, filter(
	el IN collect(
		{neighbor: suc, edge: out_edge})
	WHERE NOT el.neighbor IS NULL) as suc_maps
// match predecessors and in-edges
OPTIONAL MATCH (pred)-[in_edge:edge]->(a) 
WITH a, a1, suc_maps, filter(
	el IN collect(
		{neighbor: pred, edge: in_edge})
	WHERE NOT el.neighbor IS NULL) as pred_maps

// copy all incident edges of the original node
FOREACH (suc_map IN suc_maps |
	FOREACH(suc IN [suc_map.neighbor] |
		CREATE (a1)-[new_edge:edge]->(suc) 
		SET new_edge = suc_map.edge))
FOREACH (pred_map IN pred_maps |
	FOREACH(pred in [pred_map.neighbor] |
		CREATE (pred)-[new_edge:edge]->(a1) 
		SET new_edge = pred_map.edge))

// copy self loop
FOREACH (suc_map IN suc_maps | 
	FOREACH (self_loop IN 
		CASE WHEN suc_map.neighbor=a
		THEN [suc_map.edge] ELSE [] END |
			CREATE (a1)-[new_edge:edge]->(a1) 
			SET new_edge = self_loop))
WITH a, a1
RETURN a1
\end{lstlisting}

\vspace{20pt}
\section{Merge Cypher query}\label{App:merge}

\begin{lstlisting}

// As the following is not allowed by Cypher
// `SET a[key] = b[key]`, so we use APOC instead:
// `SET a = apoc.map.setKey(a, key, b[key])`
MATCH (a { id : 'a'}), (b { id : 'b'}) 

// Add properties of 'b' to 'a'
FOREACH(key in keys(b) |
	FOREACH(dummy IN 
		CASE WHEN key IN keys(a) 
		THEN [] ELSE [NULL] END |
			// SET a[key] = b[key]
			SET a = apoc.map.setKey(
				a, key, b[key]))
	FOREACH(dummy IN 
		CASE WHEN key IN keys(a) 
		THEN [NULL] ELSE [] END |
			SET a = apoc.map.setKey(
				a, key, a[key] + filter(
					el IN b[key] WHERE NOT el in a[key]))))
// list with ids of merged nodes to track self loops
WITH a as merged_node, b,
	[id(a), id(b)] as merged_nodes

// match successors of 'b'
OPTIONAL MATCH (b)-[out_edge:edge]->(suc)
WITH merged_node, b, merged_nodes, filter(
	el IN collect({neighbor: suc, edge: out_edge})
	WHERE NOT el.neighbor IS NULL) AS all_suc_maps
WITH merged_node, b, merged_nodes,
	filter(
		el in all_suc_maps 
		WHERE NOT id(el.neighbor) IN merged_nodes)
			AS new_suc_maps,
	filter(
		el in all_suc_maps
		WHERE id(el.neighbor) IN merged_nodes)
			AS loop_suc_maps

//  match predecessors of 'b' 
OPTIONAL MATCH (pred)-[in_edge:edge]->(b)
WITH merged_node, b, merged_nodes, new_suc_maps,
	loop_suc_maps, filter(
	el IN collect({neighbor: pred, edge: in_edge})
	WHERE NOT el.neighbor IS NULL) AS all_pred_maps
WITH merged_node, b, merged_nodes, new_suc_maps,
	loop_suc_maps, filter(
		el IN all_pred_maps
		WHERE NOT id(el.neighbor) IN merged_nodes)
			AS new_pred_maps,
	filter(
		el IN all_pred_maps
		WHERE id(el.neighbor) IN merged_nodes)
			AS loop_pred_maps

// create edges for sucs/preds that 
// didn't exist before and/or merge 
// their attributes into existing edges
FOREACH (suc_map IN new_suc_maps |
	FOREACH(suc IN [suc_map.neighbor] | 
		MERGE (merged_node)-[edge:edge]->(suc)
		// Merge dicts
		FOREACH(key in keys(suc_map.edge) |
			FOREACH(dummy IN 
				CASE WHEN key IN keys(edge)
				THEN [] ELSE [NULL] END |
				// SET edge[key] = suc_map.edge[key]
				SET edge = apoc.map.setKey(
					edge, key, suc_map.edge[key])
			)
			FOREACH(dummy IN 
				CASE WHEN key IN keys(edge) 
				THEN [NULL] ELSE [] END |
					SET edge = apoc.map.setKey(
						edge, key, edge[key] + filter(
							el IN suc_map.edge[key] 
							WHERE NOT el in edge[key]))))))

FOREACH (pred_map IN new_pred_maps |
	FOREACH(pred in [pred_map.neighbor] |
		MERGE (pred)-[edge:edge]->(merged_node)
		FOREACH(key in keys(pred_map.edge) |
			FOREACH(dummy IN
				CASE WHEN key IN keys(edge)
				THEN []
				ELSE [NULL] END |
					SET edge = apoc.map.setKey(
						edge, key, pred_map.edge[key])
			)
			FOREACH(dummy IN
				CASE WHEN key IN keys(edge)
				THEN [NULL] ELSE [] END |
				SET edge = apoc.map.setKey(
					edge, key, edge[key] + filter(
						el IN pred_map.edge[key]
						WHERE NOT el in edge[key]))))))

// handle self loops
WITH merged_node, b, loop_suc_maps, loop_pred_maps
OPTIONAL MATCH (merged_node)-[old_loop:edge]->(merged_node)
FOREACH(dummy IN 
	CASE WHEN NOT old_loop IS NULL OR
		length(loop_suc_maps) > 0 OR
		length(loop_pred_maps) > 0
	THEN [NULL] ELSE [] END |
	MERGE (merged_node)-[
			old_loop:edge]->(merged_node)
	FOREACH (map IN loop_suc_maps + loop_pred_maps |
		FOREACH(key in keys(map.edge) |
			FOREACH(dummy IN
				CASE WHEN key IN keys(old_loop)
				THEN [] ELSE [NULL] END |
				SET old_loop = apoc.map.setKey(
					old_loop, key, map.edge[key])
			)
			FOREACH(dummy IN
				CASE WHEN key IN keys(old_loop)
				THEN [NULL] ELSE [] END |
				SET old_loop = apoc.map.setKey(
					old_loop, key, old_loop[key] + filter(
						el IN map.edge[key]
						WHERE NOT el in old_loop[key]))))))
DETACH DELETE b
RETURN merged_node
\end{lstlisting}

\end{document}